\documentclass[a4paper,11pt]{article} 
\usepackage{psfig}
\usepackage{glas}  
\newcommand{\alphas}{$\alpha_s$} 
\begin{document} 
 
 
 \begin{titlepage}{GLAS-PPE/2006-16}{24$^{\underline{\rm{th}}}$
September 2006}
\title{QCD PHYSICS WITH ZEUS AND H1 AT HERA} 
 
\author{P. J. BUSSEY\\ 
Department of Physics and Astronomy, University of Glasgow,\\ 
Glasgow G12 8QQ, U.K.\\ 
p.bussey@physics.gla.ac.uk} 
 
\maketitle 
 
 
\begin{abstract} 
A review is presented of recent results in QCD from  
the H1 and ZEUS experiments at HERA, emphasizing the use of  
higher order calculations to describe the data. 
Keywords: {QCD, Quantum Chromodynamics, H1, ZEUS, electron, positron, proton, photon, jet, quark, gluon, collisions} 
\end{abstract} 
 
PACS Nos.: 12.38.Qk, 13.60.Hb, 13.87.Ce, 14.20.Dh, 14.65.-q, 14.70.Bh
 
\end{titlepage}

\section{Introduction}	 
 
The theory of Quantum Chromodynamics (QCD) is a central component of
the Standard Model of elementary particle physics.  It accounts for
the strong nuclear force in terms of the exchange of spin-one entities
known as gluons; these carry a quantum number known as ``colour'' and are 
exchanged between themselves and between the fermions referred to as
quarks, which carry electric as well as colour charge.  There is a law of
nature that the overall colour of a stable elementary particle must be
zero.  Thus the quarks and gluons are never observed alone, but
mesons are formed (primarily) from quark-antiquark pairs and baryons
from three quarks or antiquarks.  The quarks that provide the basic
constitution and quantum numbers of a hadronic particle are known as
valence quarks.
 
When a high-energy electron (or positron) collides with a proton, as
in the HERA collider, the simplest nuclear reaction that
can occur is that a quark is ejected to give a high-energy jet of
particles.  This is Deep Inelastic Scattering (DIS) and is mediated by
the exchange, between the electron and the quark, of a virtual photon,
$W$ or $Z$. Scattering processes can involve valence quarks, but can
also occur by means of virtual quark-antiquark pairs that appear
temporarily within the proton. Indeed, the proton may be considered as
consisting of a continuum of different combinations of partons, that
is to say (anti)quarks and gluons, and depending on how violently it is
struck by the exchanged boson. Thus the parton density functions
(PDFs) of the proton are functions both of the fraction $x$ of the protons
taken by a given parton and of the virtuality $Q^2$ of the
exchanged boson.
 
\section{The HERA Collider} 
The HERA Collider has been operating at DESY, Hamburg, since the year
1992.  Electrons or positrons are stored at 27.5 GeV and collide with
a stored proton beam of energy 920 GeV. The experiments H1 and ZEUS
are located at the collision points and record the various $ep$
interactions that occur. Over the years, the collider has achieved
increasingly high luminosities which have enabled the different types
of process to be studied in greater depth and detail. During 2001-2002
an upgrade to the collider and experiments was carried out; however,
earlier analyses have continued and many of the results presented here
come from these.  Using silicon and drift chamber systems, each of the
experiments is able to record tracks of charged particles emanating
from the interaction point, supplemented by electromagnetic and
hadronic calorimeters to measure the overall energy of jets and high
energy single particles.  The apparatus is asymmetric, with
calorimetry in the proton (forward, ``positive'') direction designed
to record large hadronic energy deposition, and calorimetry in the
electron (``negative'') direction aimed at measuring precisely the
scattered electron.  Both electron and positron beams were used.
 
\section{Tests of QCD at HERA} 
Tests of QCD at HERA may broadly be considered as of the following kinds: 
\begin{itemize} 
\item The PDF's of the proton measured at different $Q^2$ values should be  
consistent with each other, as described by the DGLAP equations. 
\item Cross sections for processes with different numbers of QCD vertices should be predictable.  Here, in particular, we are looking at processes with
different numbers of high transverse momentum ($p_T$) jets, many of
which correspond to radiated gluons, each of which represents an
additional QCD vertex.  We note that the simple lowest-order (LO) DIS
process has no vertices involving gluons, and is purely
electromagnetic or electroweak in nature once the PDF is known.
\item The value of the strong coupling constant \alphas, which acts at quark-gluon 
and gluon-gluon vertices, must be consistent when measured in
different ways.  The value of \alphas\ varies (``runs'') with the
momentum exchanged at the relevant vertex, in a way specified by QCD theory.

\end{itemize} 
 
Various types of process are useful for these studies.  Owing to the
complexity of the QCD calculations, the processes that can be
calculated theoretically are those that involve perturbative
expansions in terms of \alphas, starting at LO. Even next-to-leading
order calculations (NLO) are highly complex, however.  Many such
calculations have now been completed, and a few NNLO calculations,
with special methods (resummation) employed to take account of
multiple initial-state gluon radiation from the proton system.  Since
the running value of \alphas\ falls with the momentum exchanged in the
collision process, high values of exchanged momentum are preferred,
so-called ``hard'' processes.  ``Soft'' processes require a
higher-order or non-perturbative approach which is not usually
achievable from first principles.  The following types of hard process
have been studied particularly at HERA; more than one of these
features may of course be present in a given reaction.
\begin{itemize} 
\item DIS processes, i.e.~those involving high $Q^2$ of the exchanged boson. 
\item Processes involving high-$p_T$ jets. 
\item The production of heavy-quark systems, i.e.~charm and beauty quarks. 
\item Hard diffractive processes in which the exchanged colour-neutral
object may be treated partonically.
\end{itemize}

\begin{figure}[t]
~\\[-5mm]
\centerline{\mbox{
\psfig{file=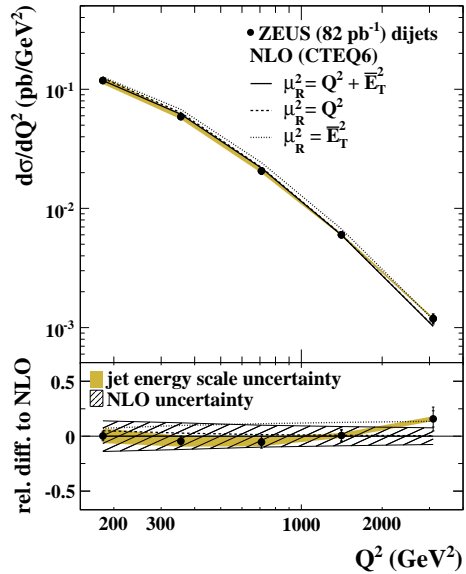,width=2.6in}
}}
\caption{Inclusive dijet production in DIS from ZEUS, with NLO QCD fit.}
\label{ZDISinjet06}
\end{figure}

\section{Jets in DIS}
A long-standing analysis program at HERA has involved studying the
PDF's of the proton by means of the ejection of one or more jets by the
incident virtual boson.  At lowest order (LO) only a single parton is
involved, and at a fixed $Q^2$ value the process does not display
explicit QCD features.  However the variation of the PDF's with $Q^2$
is given by QCD-governed evolution equations (the DGLAP equations).
These have been well verified in processes dominated by $u$ and $d$ type
quarks.  When inclusive jets are measured, the contributions from 
multi-jet final states are included, and here QCD-governed 
processes contribute.  Both H1 and ZEUS have published QCD-based analyses
of the parton structure of the proton.\cite{HZPDF} 

The two experiments have recently measured inclusive jet production to
improved precision in high-$Q^2$ neutral-current DIS.  It is
convenient to perform DIS jet measurements in the Breit frame, in
which the incoming virtual boson and proton define the event axis.  If
the outgoing state contains the proton remnant and just one jet, the
latter then has no transverse momentum relative to this axis.
High-$E_T$ jets in the Breit Frame therefore indicate a $\ge2$-jet
process which is sensitive to QCD physics.  Dijet measurements by ZEUS
are shown in Fig.\ \ref{ZDISinjet06} for a range of exchanged boson
virtualities.\cite{Zincjet} The jet cross sections have been evaluated
as functions of several kinematic and jet variables and are well
described by the NLO QCD calculation DISENT.\cite{DISENT} The aim has
been to identify regions of phase space where uncertainties from
higher-order corrections are small, so as to facilitate the evaluation
of improved gluon densities in the proton in global fits to the proton
PDFs.

\begin{figure}[t]
\centerline{
\raisebox{1mm}{\psfig{file=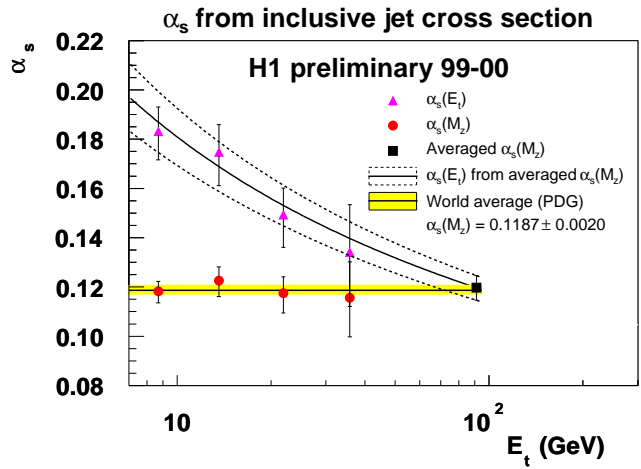,width=2.4in,angle=-90}}
}
\caption{Results of fit for \alphas\  by H1 to inclusive jet data, as a function of jet transverse energy.}
\label{Hinjet06}
\end{figure}
H1 have used the NLO QCD program NLOJET++ to represent their data at
the parton level.\cite{NLOJET} It is performed in the
$\overline{\mathrm{MS}}$ scheme with five flavours, the effects of
hadronisation being evaluated using the generators 
RAPGAP (based on LO matrix elements) and DJANGOH (using the Colour Dipole
Model).\cite{RAPGAP,DJANGOH} Used together with $Q^2$-varying
PDFs, this gives a good description of the data and allows a value of
$\alpha_s$ to be extracted by varying $\alpha_s$ in the fits. 
The fitted $\alpha_s$ values (Fig.\ \ref{Hinjet06}) show running, with
the equivalent value at $M_Z$ showing good consistency as a function of
transverse jet energy. A value of $0.01197\pm0.0016\pm0.0047$ is
obtained, in agreement with other determinations and with
competitive precision.

\begin{figure}[t]

\centerline{
\mbox{\hspace*{-2mm}
\psfig{file=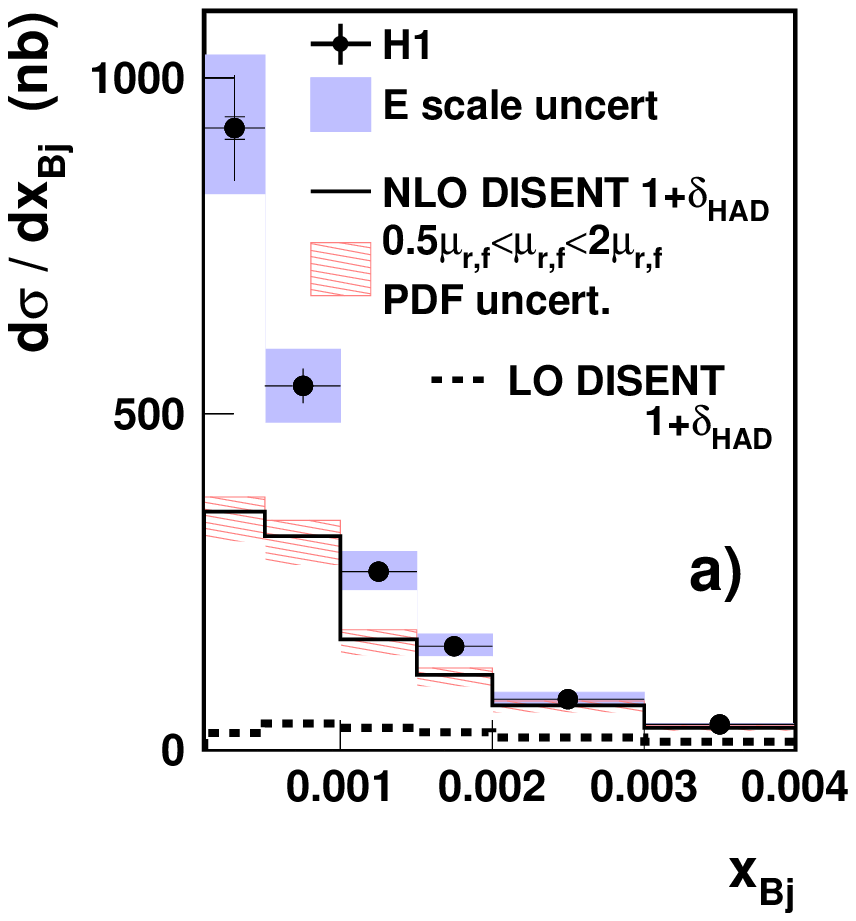,width=1.8in}
\hspace*{-7mm}
\psfig{file=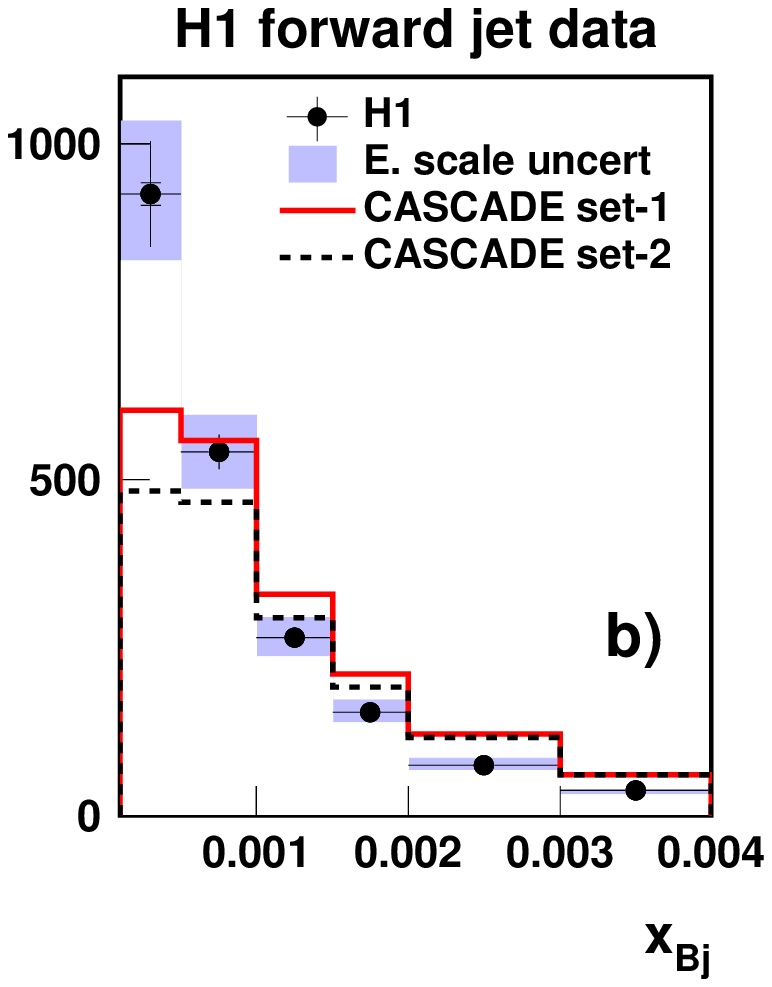,width=1.8in}
\hspace*{-7mm}
\psfig{file=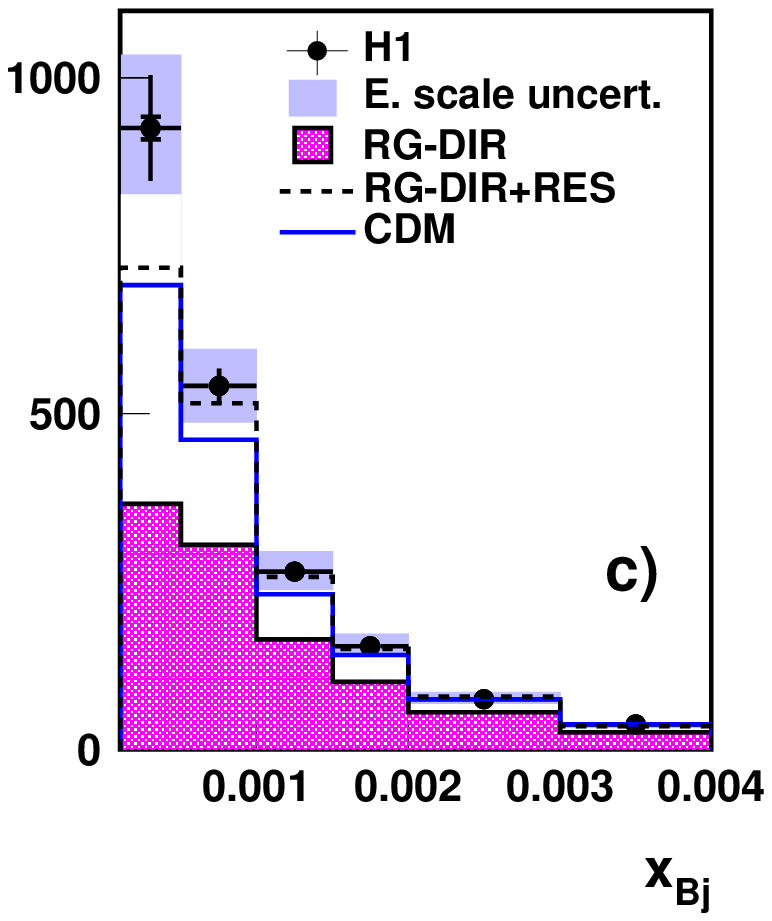,width=1.8in}
}}
\caption{H1 measurements of jet production cross sections 
at low $x$ in DIS  compared to various theoretical models (see text). 
}\label{H1forj}
\end{figure}

H1 and ZEUS have measured the inclusive production of jets in DIS in
the forward direction.  Using DISENT,
H1 find that their cross sections (Fig. \ref{H1forj}) are not well
described by LO or NLO calculations in
QCD.\cite{H1forjx} Models with a resolved photon contribution, and the
Colour Dipole Model (CDM) implemented in ARIADNE, give an
improvement,\cite{ARIADNE} as does the CASCADE model which
incorporates the CCFM description of the proton using
$k_T$-unintegrated parton densities.\cite{CASCADE} However none of the
models considered is satisfactory, and H1 conclude that a higher-order
description is required.

\begin{figure}[t]
~\\[-10mm]
\centerline{
\hspace*{5mm}\psfig{file=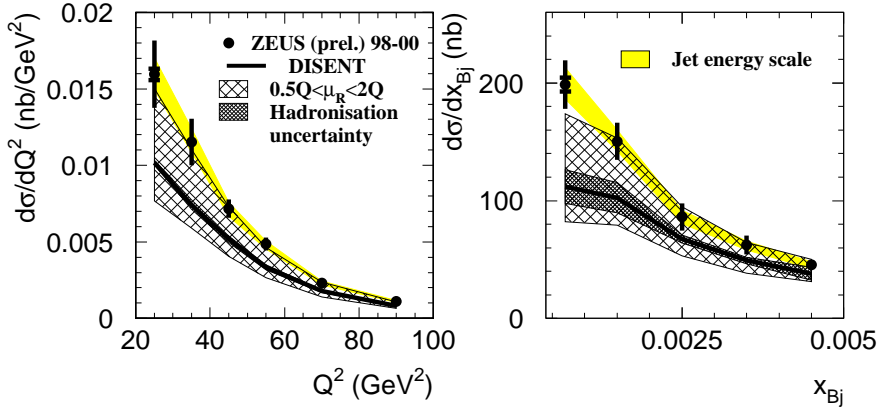,width=5.0in}\\[-6mm]
}
\caption{Preliminary ZEUS measurements of forward jet production cross 
sections at low $x$ in DIS compared to the DISENT model.
}\label{Zforj}
\end{figure}
ZEUS however, without going to the very lowest
$x$ values attained by H1, find that DISENT is satisfactory (Fig. \ref{Zforj})
in an update of a previous analysis.\cite{Zfjet0}
They assign large values to the 
QCD scale uncertainties in the perturbative calculation, and these
cover the discrepancies between the latter and the data.
It is evident that there is a theoretical issue here which needs to be
resolved, and as yet these measurements
do not give clear evidence in favour of the CCFM model. 

\begin{figure}[t]
~\\[-3mm]
\centerline{
\mbox{
\psfig{file=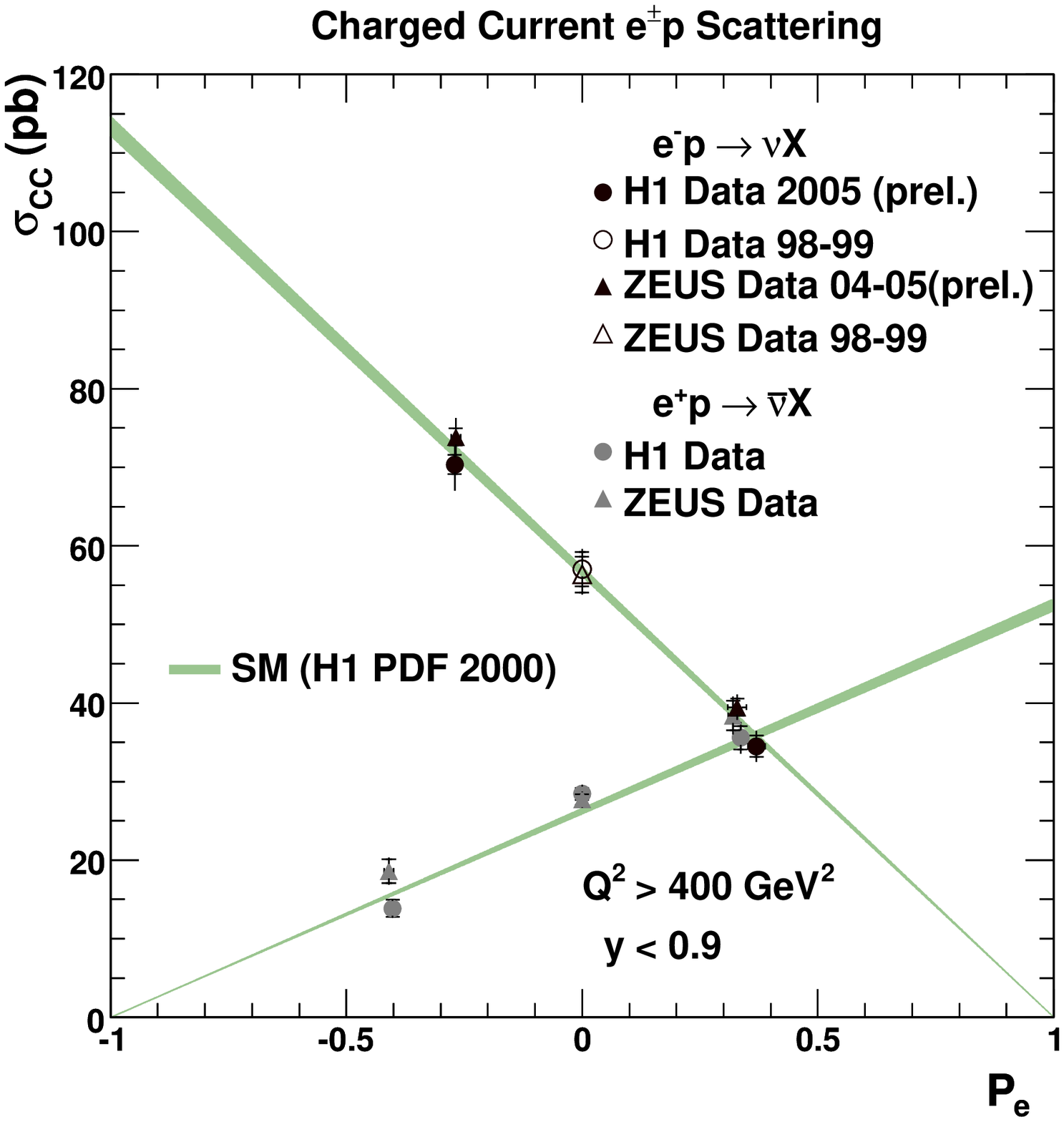,width=2.6in}
\hspace*{-5mm}
\raisebox{-1mm}{\psfig{file=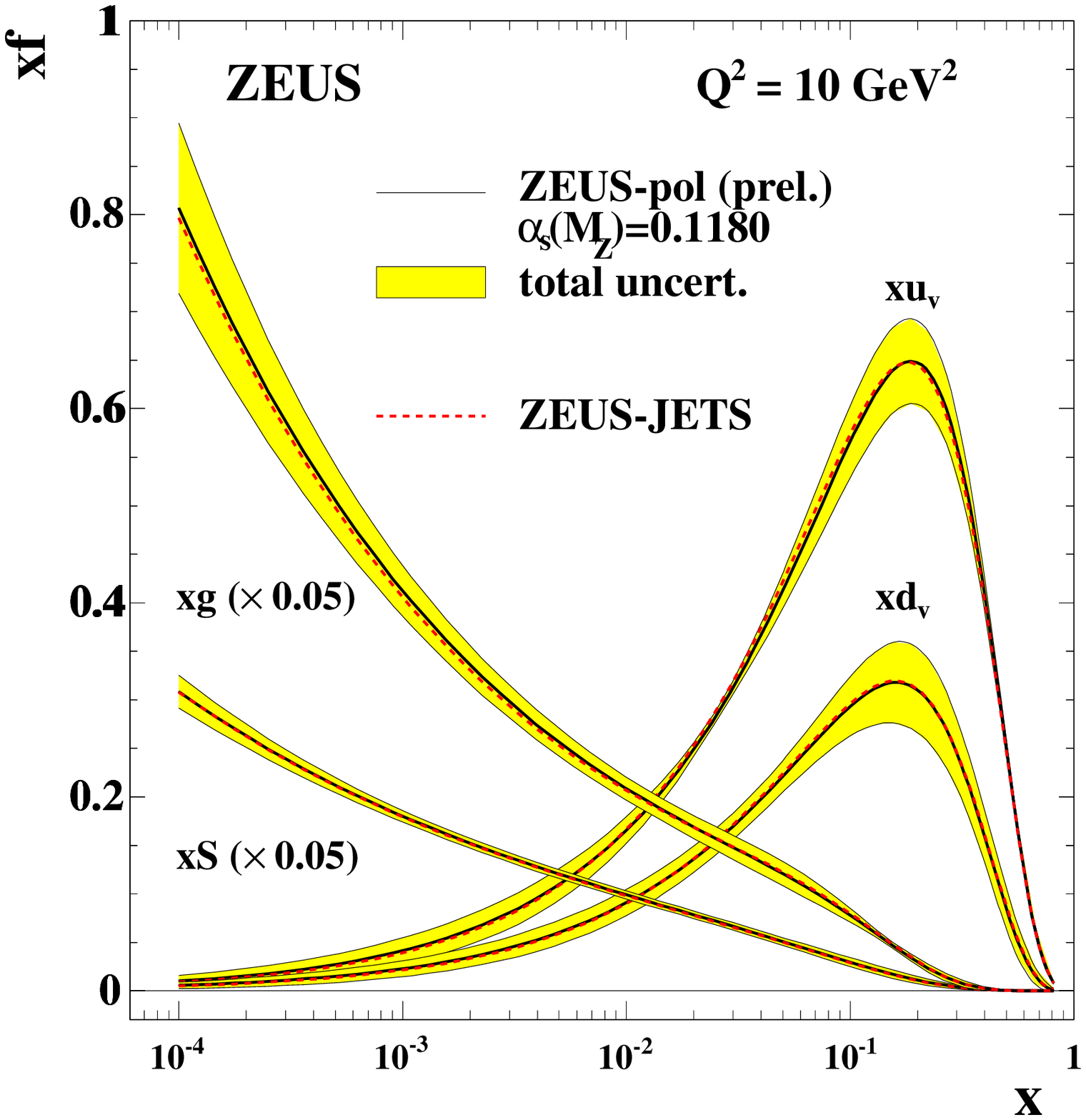,width=2.56in}}
}}
\caption{(Left) charged-current scattering cross sections of electrons 
(black) and positrons (grey) from ZEUS and H1 as a function of lepton
polarization, compared to standard model predictions. (Right) Parton
densities from the ZEUS fits to jet data (ZEUS-JETS), and
incorporating the new polarisation data (ZEUS-pol); the uncertainties
are those on the ZEUS-pol fits.}\label{pola}
\end{figure}
At high values of $Q^2$, cross sections have been measured for
charged-current DIS using the newly developed polarised $e^\pm$ beams
at HERA (Fig. \ref{pola}).  The differences between the two cross
sections as a function of polarisation are very evident and in
agreement with the Standard Model, using the Monte Carlos ARIADNE and
MEPS to describe the hadronisation effects.  The earlier ZEUS NLO PDF
analysis has been updated to include these data, and the resulting
picture is entirely consistent.

\section{Jet structure and event shapes}
\begin{figure}[t]
~\\[-6mm]
\centerline{
\mbox{\hspace*{-3mm}
\psfig{file=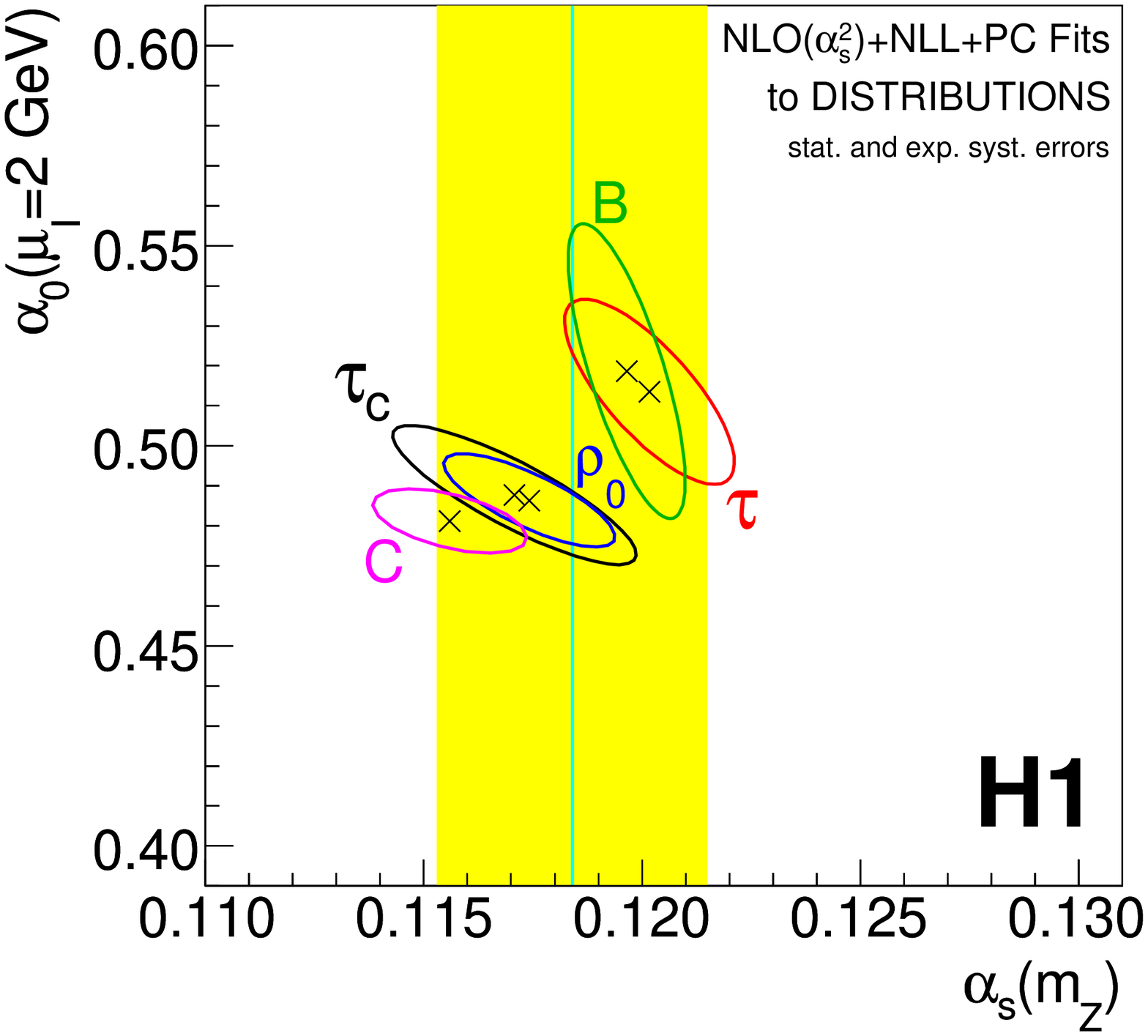,width=2.66in}
\raisebox{2mm}{\psfig{file=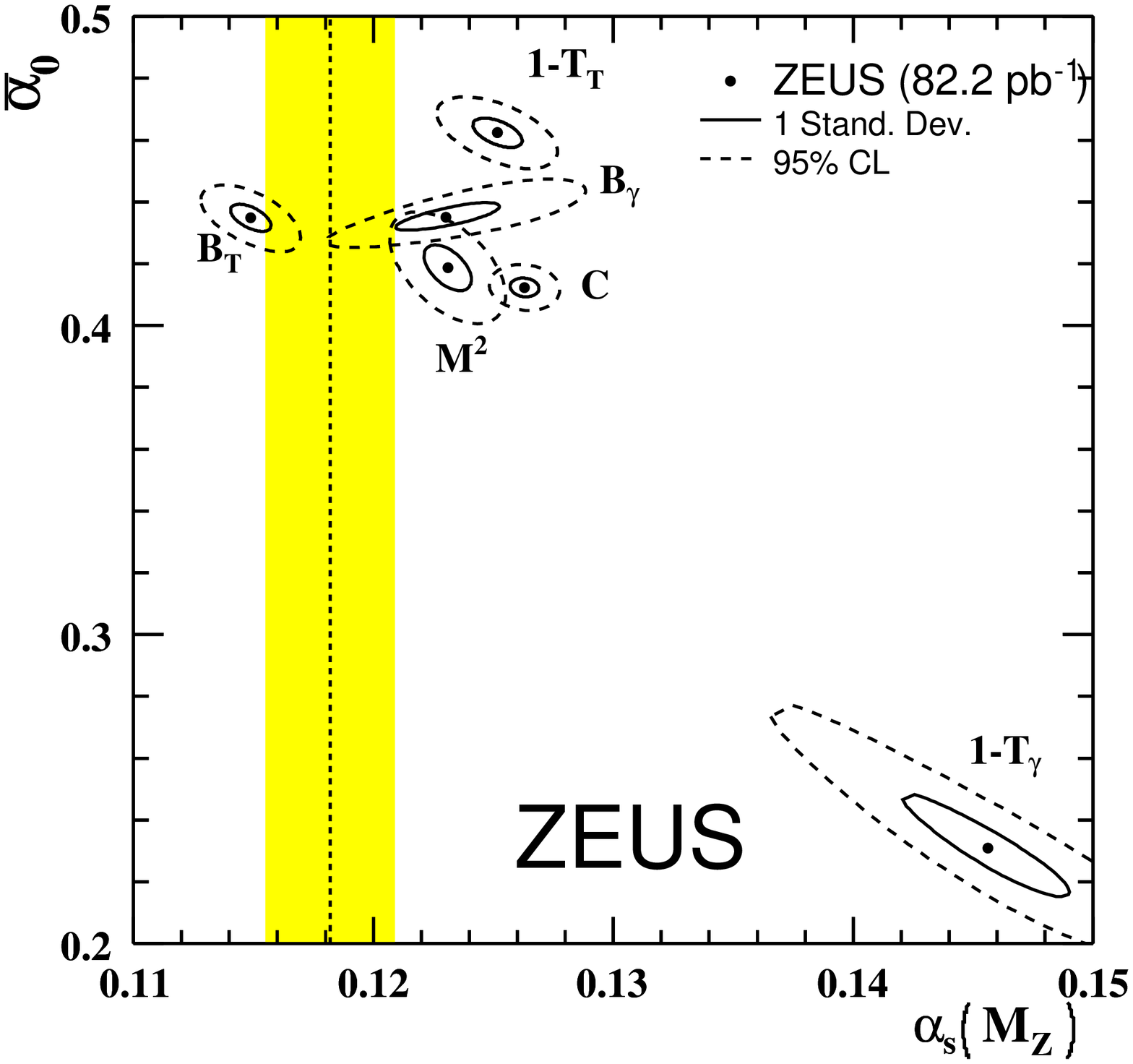,width=2.4in}}
}}
\caption{Measurements by H1 and ZEUS of event shape variables:
$\tau,\,T$ = thrust; $c$-parameter; $B$ = broadening; $M^2, \rho_0$
=invariant jet mass.  Subscripts $T,\, C$ denote the thrust axis (ZEUS,
H1), $\gamma$ the incoming photon axis (ZEUS); unsubscripted $\tau$
and $B$ (H1) are relative to the photon axis.  }\label{evsh}
\end{figure}

The particles in a jet are produced by the soft fragmentation of the
initiating high-$p_T$ parton.  The QCD processes involved here are
highly non-perturbative and must be modelled phenomenologically, as is
done in the leading-logarithm Monte Carlos PYTHIA and HERWIG.
However, harder radiation of partons within the jet can deform its
shape from a long pencil-like object (with high {\it thrust\/}) into a
form with broader and asymmetric characteristics.  These can be
described by various event-shape variables, whose values are given by
an $\alpha_s$-dependent term together with so-called power corrections
to account for the soft effects and ensure that these are suitably
merged on to the hard processes.\cite{powerc} Both ZEUS and H1 have
extended their measurements of a variety of event-shape
variables,\cite{H1evshp,Zevshp} and have extracted values of
$\alpha_s$ together with a parameter $\alpha_0$ which, according to
power-correction theory, should be universal with a value of
$\approx0.5$ (Fig. \ref{evsh}).  Differential distributions in the
event shape variables have also been studied. The theoretical
expectations are reasonably well observed in most of the event shapes
measured, but there is a lack of consistency in the extracted
$\alpha_0$ and $\alpha_s$ variables that may indicate a need for
higher order terms.  The two experiments also disagree over the
detailed values extracted from their fits, apparently because
different kinematic regions were used. This gives an indication of the
systematic uncertainties currently involved with the power correction
method.

\begin{figure}[t]
~\\[-10mm]
\centerline{
\mbox{
\psfig{file=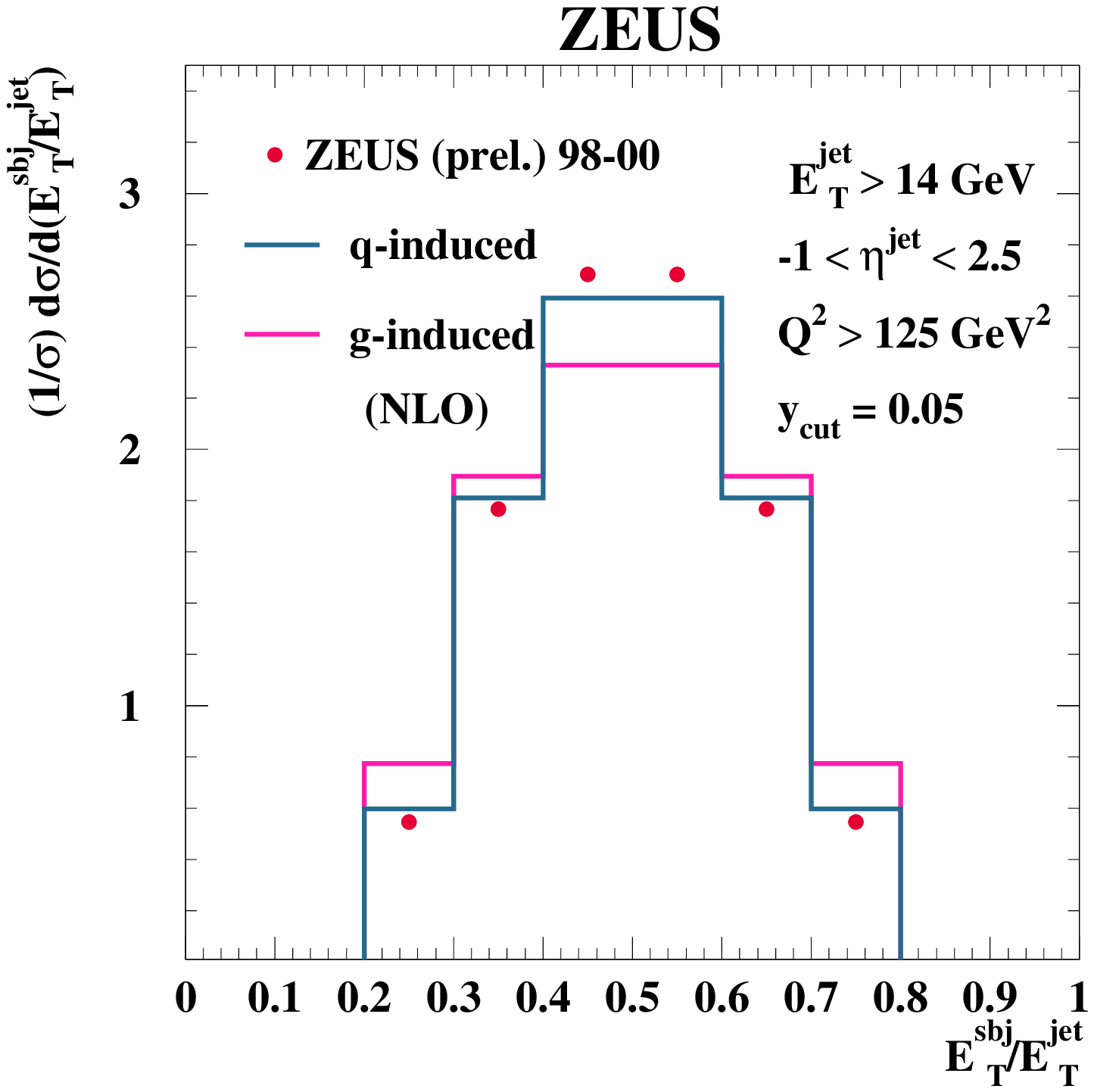,width=3.1in}
\hspace*{-14mm}
\psfig{file=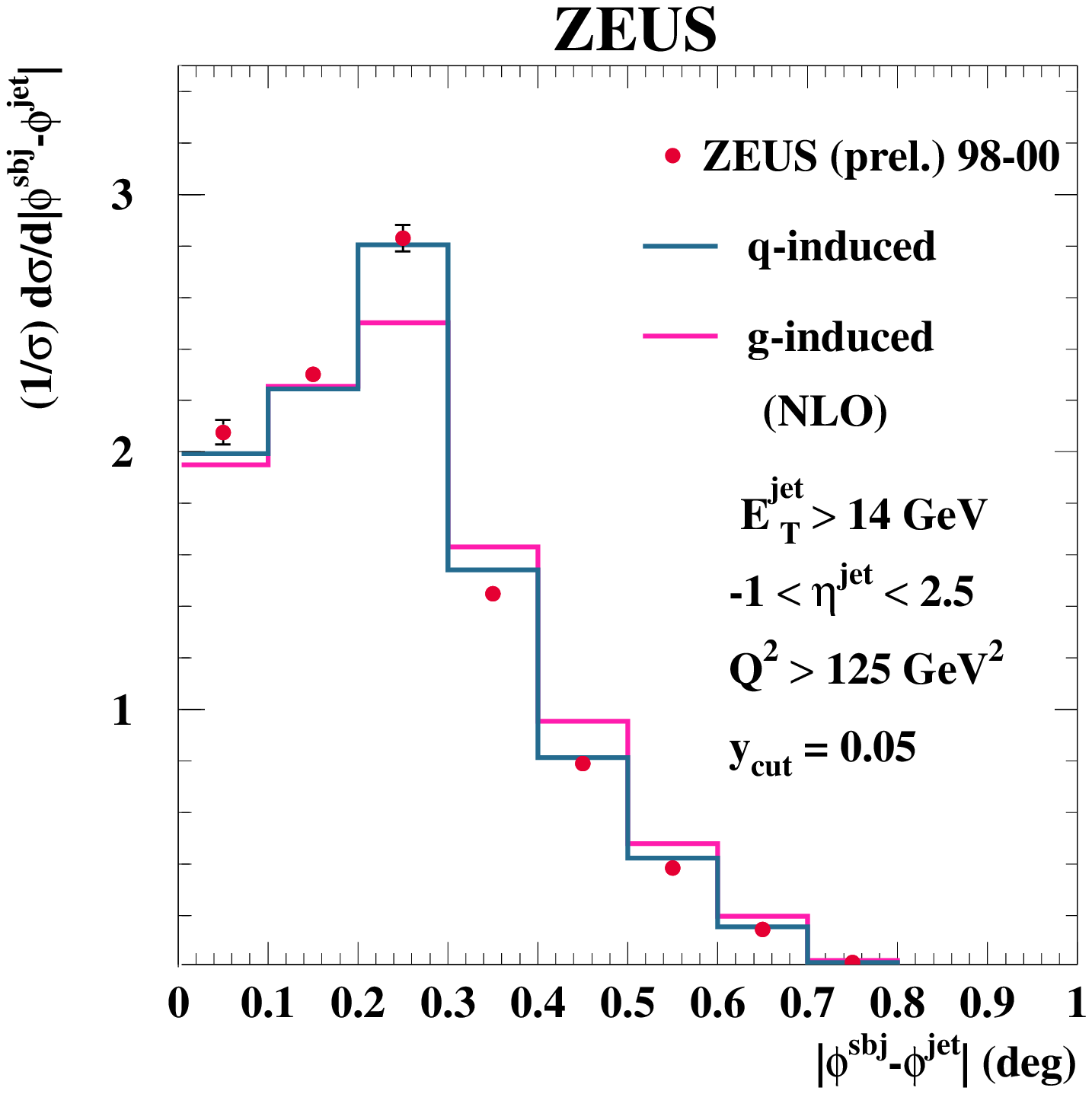,width=3.1in}}\\[-12mm]
}
\caption{Kinematic properties of subjets in DIS from ZEUS: energy fraction (left) and pseudorapidity difference (right), compared to NLO prediction using DISENT.}
\label{subjets}
\end{figure}
A further probe of the hard structure of QCD jets comes from the study
of subjets.  These are defined as jet-like substructures within jets,
which may be obtained by reapplying the jet cluster algorithm at a
smaller resolution scale $y_{cut}$ than was used for identifying the
jet itself.  Subjet have been studied by ZEUS in neutral current DIS.
The jets were identified using the $k_T$ cluster algorithm in the
longitudinally invariant inclusive mode in the laboratory frame.
Measurements of subjet distributions were carried out as functions of
various kinematic quantities, including the ratio of the subjet
transverse energy to that of the jet, and the differences between the
pseudorapidities and the azimuth values of the subjet and of the
jet. The measured distributions are used to study the pattern of
parton radiation by comparing them with Monte Carlo models and
perturbative QCD calculations. It is found that ARIADNE performs
better than LEPTO-MEPS and that comparison with NLO predictions allows
a confirmation that quark rather than gluon jets are mostly being
observed (Fig.\ \ref{subjets}).

\section{Multi-jet properties} 
\begin{figure}[t]
~\\[-1mm]
\centerline{
\mbox{
\psfig{file=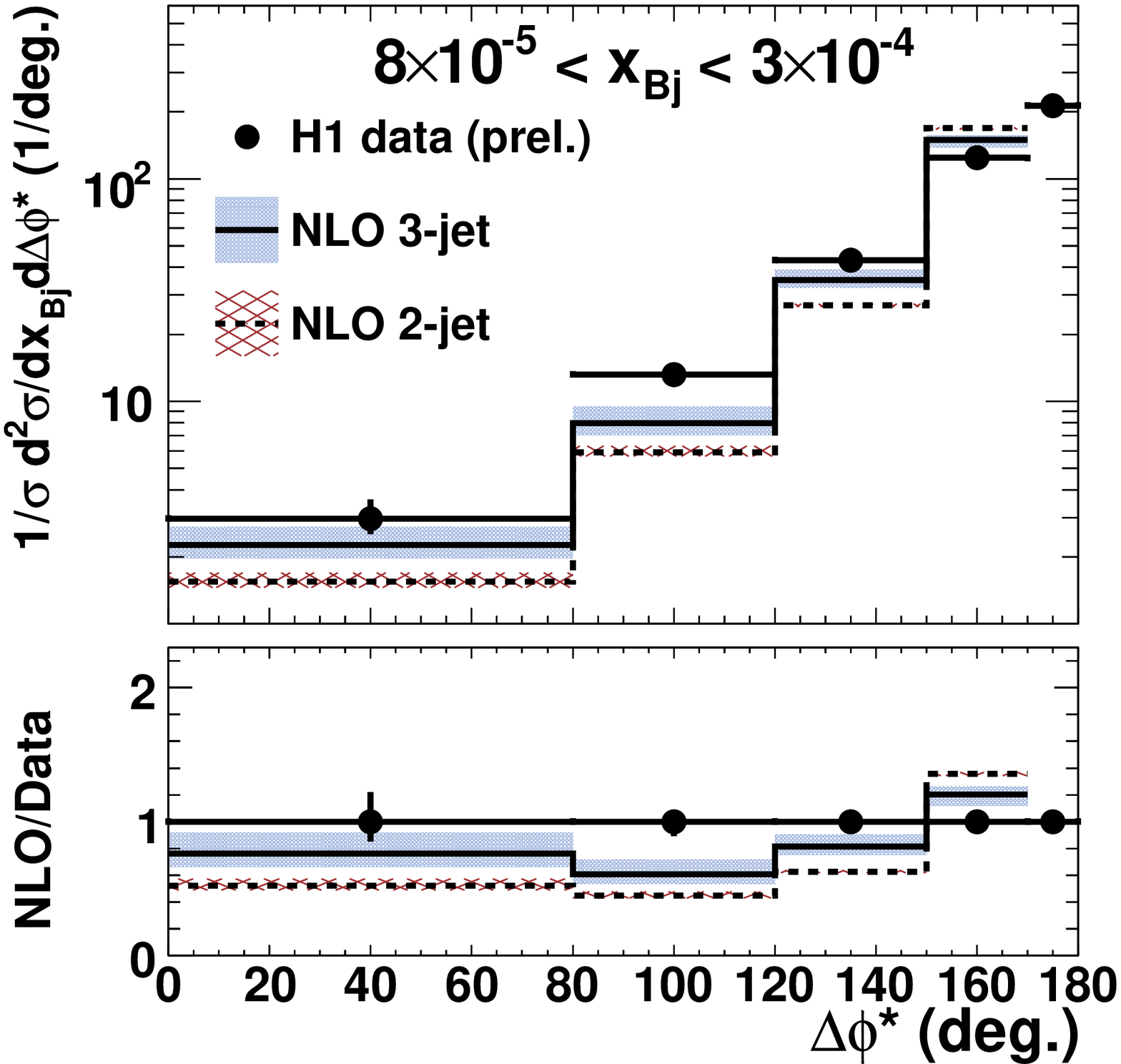,width=2.1in}
\hspace*{3mm}
\psfig{file=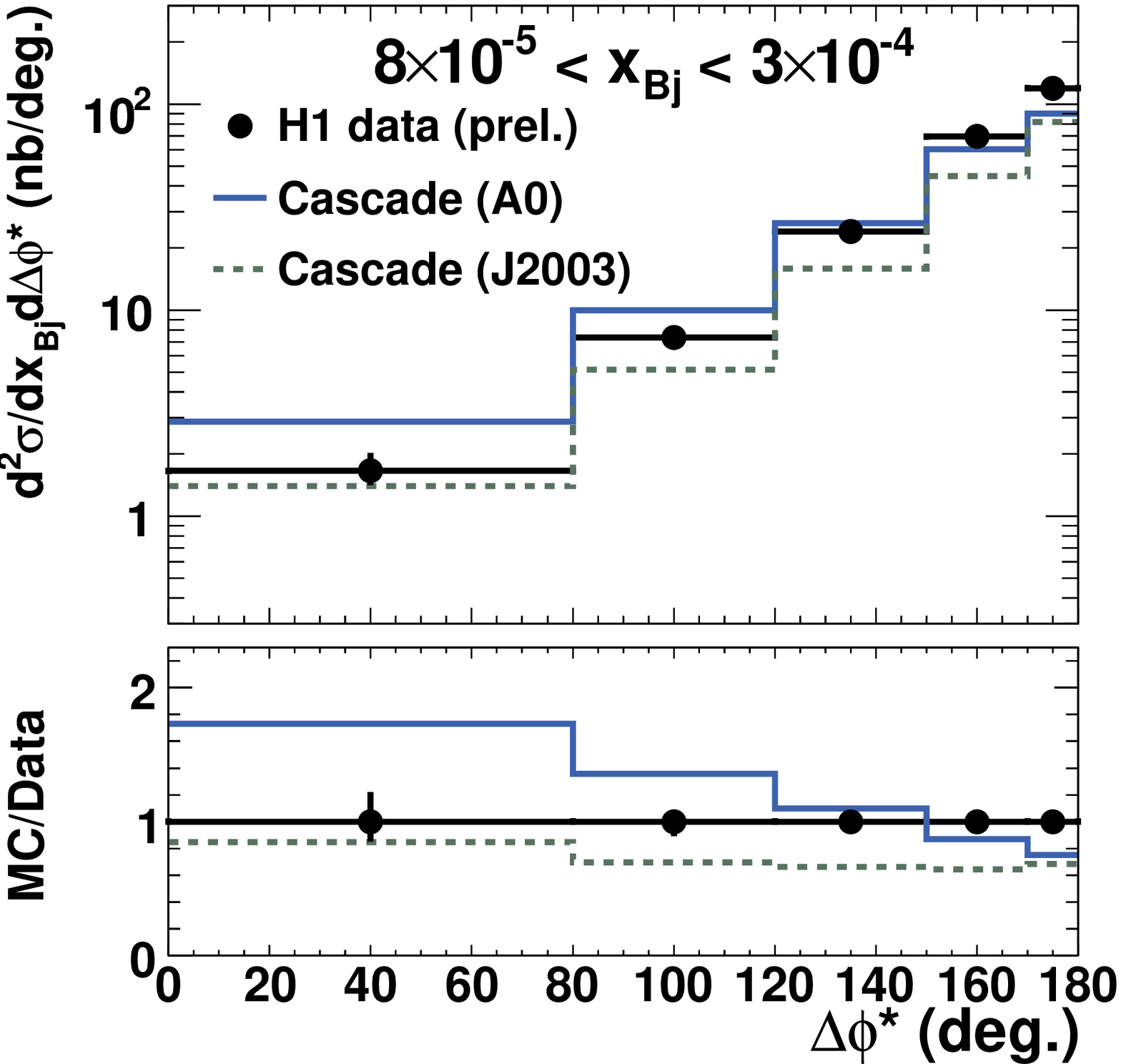,width=2.1in}
}}
\caption{Azimuthal correlations between jets in two-jet 
final states in DIS measured in the hadronic centre-of-mass frame.  
The NLO predictions are normalised to the data outside the highest $\Delta\phi^*$ bin, while the CASCADE predictions are absolute.}\label{H1dijaz}
\end{figure}

When two or more jets emerge in a high energy process, 
a QCD-based description is essential even at lowest order.  H1 have
measured the correlations in the azimuthal angle when measuring
two-jet final states in DIS. This distribution is sensitive to
additional parton radiation both in the initial state and from the
outgoing hard partons. H1 find that the process is sensitive to higher
order radiative processes and the simple NLO two-jet amplitudes are
insufficient to describe the data. NLOJET three-jet calculations give
improvement but are still insufficient (Fig. \ref{H1dijaz}(a)), as are
attempts to incorporate a resolved-photon component in the RAPGAP
model. The CASCADE model also fails, suggesting a need for more
theoretical input.  In principle, data of this kind can give
information on the $k_T$-unintegrated parton densities within the proton.

Similar measurements from ZEUS, however, indicate good agreement
with NLO QCD calculations for the dependence of the overall cross sections
on kinematic variables such as $Q^2$, $x$ and jet $E_T$.\cite{Zaz}  Like
H1, ZEUS require higher order terms in order to describe the azimuthal
correlations.  The inclusion of O($\alpha_s^3$) contributions increases
the predictions by an order of magnitude when the two highest-$E_T$
jets are not opposite in azimuth. Overall, the description of the 
two- and three-jet processes is reasonable although not perfect
at the highest jet-$E_T$ values. 


\begin{figure}[t]
\hspace*{-4mm}
\centerline{
\mbox{
\psfig{file=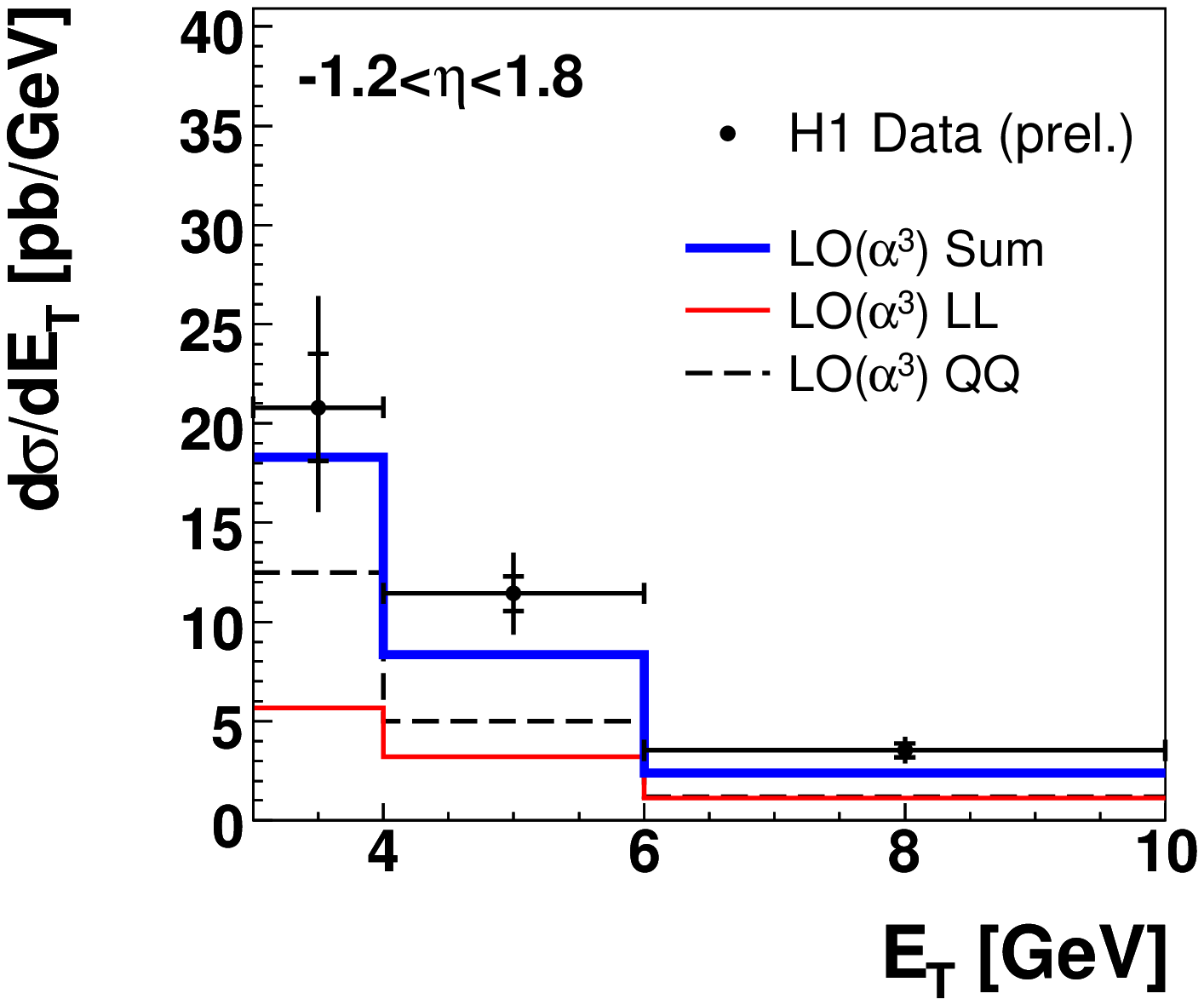,width=2.45in}
\hspace*{3mm}
\psfig{file=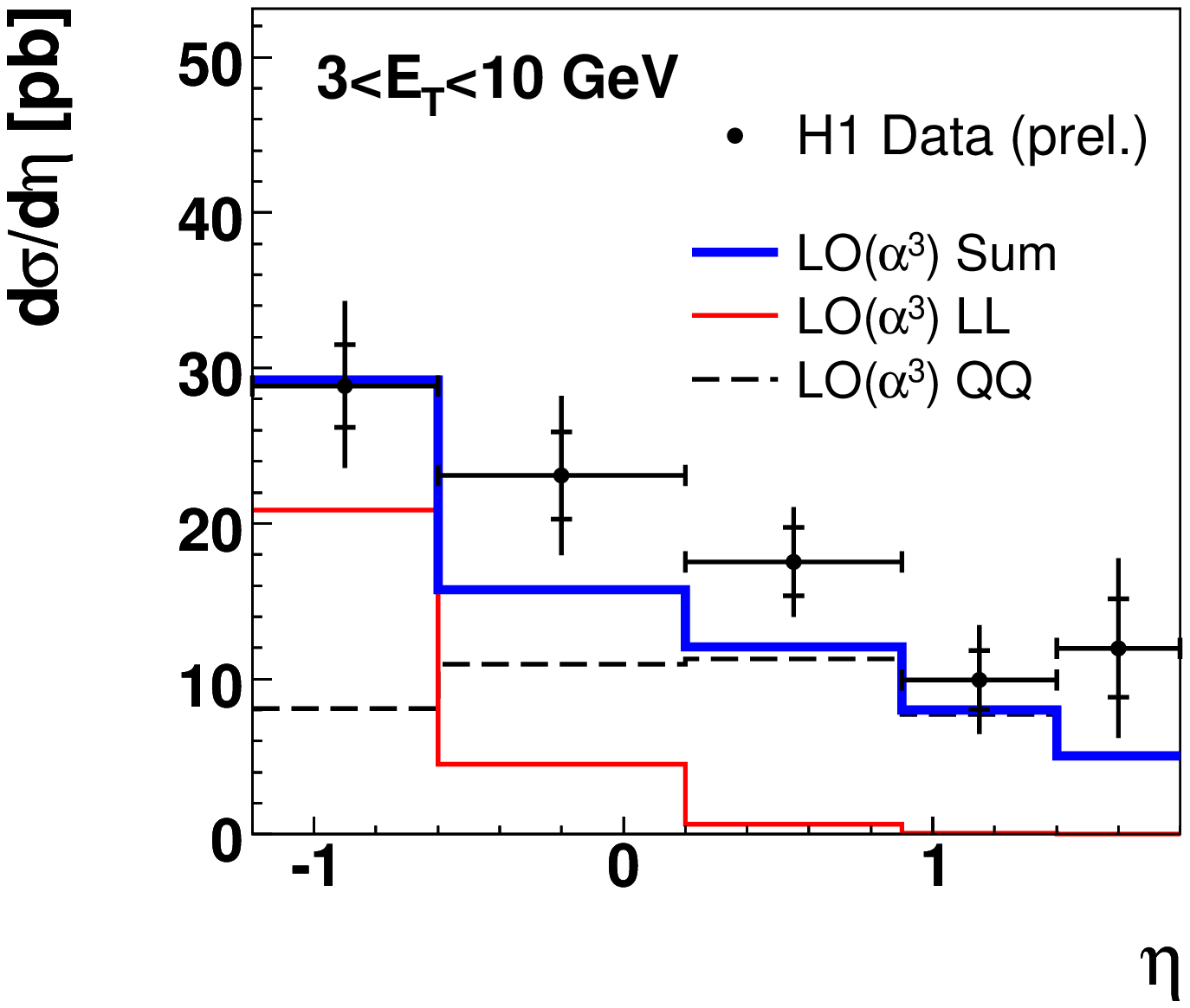,width=2.45in}
}}S
\caption{Distributions of prompt photons in DIS from H1, as a function
of photon transverse energy and pseudorapidity. The theoretical 
curves LL and QQ show the contributions of radiation off
the electron and the quark line respectively, the interference term being
small.
}\label{H1prp}
\end{figure}

\begin{figure}[t]
~\\[-10mm]
\centerline{
\mbox{
\raisebox{4mm}{\psfig{file=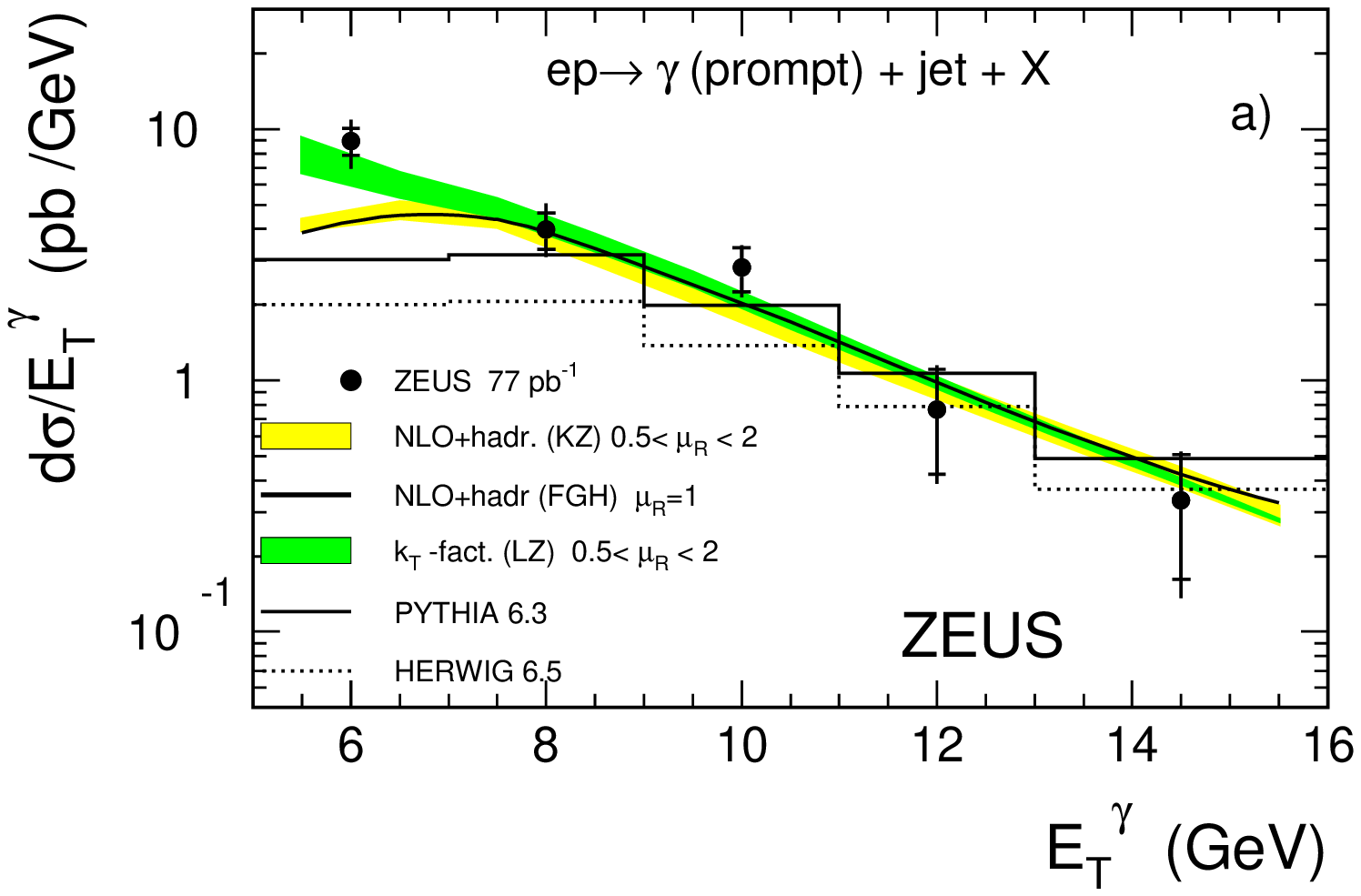,width=2.7in}}
\hspace*{-6mm}
\raisebox{0.5mm}{\psfig{file=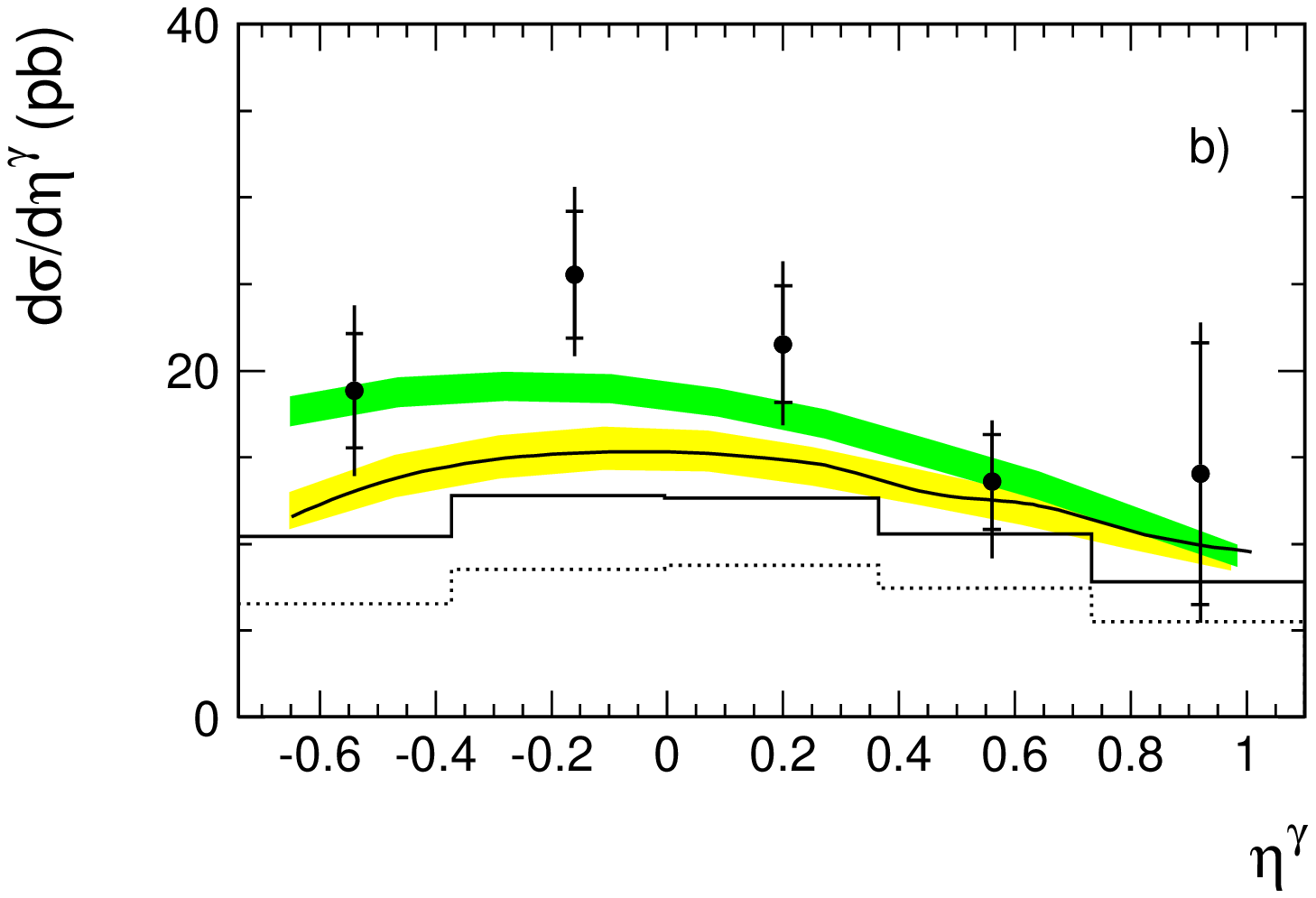,width=2.7in}}
}\\[-8mm]}
\caption{ZEUS prompt photon measurements in conjunction with a jet:
a) differential photon cross sections as a function of transverse energy,
b) as a function of laboratory pseudorapidity.}\label{Zprp}
\end{figure}
\section{Prompt photon production}
Prompt photons are those that emerge directly from a hard scattering
process, as opposed to photons found as products of decay
processes or as soft radiation from an outgoing particle.  They can
give information about QCD processes in a more direct way than do jets, which 
require hadronisation effects to be allowed for in order to be interpreted
as a measure of quark and gluon properties. 

Following a first publication by ZEUS on prompt photons in DIS, H1
have presented differential DIS cross sections.\cite{H1prp} They are
obtained for photon transverse energies of 3 - 10 GeV and laboratory
pseudorapidities of $-1.2$ - 1.8, and the data are compared with the
predictions of a new LO calculation\cite{GDR}, which gives reasonable
agreement. Comparisons with the predictions of the event generators
PYTHIA and HERWIG are also presented.  Fig.\ \ref{H1prp} illustrates
these results.

In photoproduction, the production of prompt photons with an
accompanying jet has been studied by ZEUS,
using a different technique from their earlier results.\cite{Zprp}  In the 
present case the photons were identified from their conversion probabilities
in a preshower detector installed in front of the calorimeter.
The differential $\gamma$ + jet cross sections were
reconstructed as functions of the transverse energy, pseudorapidity
and $x_\gamma$, the fraction of the incoming photon momentum taken by
the photon-jet system. Predictions based on leading-logarithm
parton-shower Monte Carlo models 
were compared with the data (Fig. \ref{Zprp}) and were found
generally to underestimate the cross sections for 
photons of transverse energy below 7 GeV, while the NLO QCD
calculations (applying the jet algorithm to the partons) 
agree with the data better.  For higher transverse
energies of the prompt photons both types of calculation 
are in good agreement with the data.

\section{Jets in photoproduction}

\begin{figure}[t]
\centerline{\psfig{file=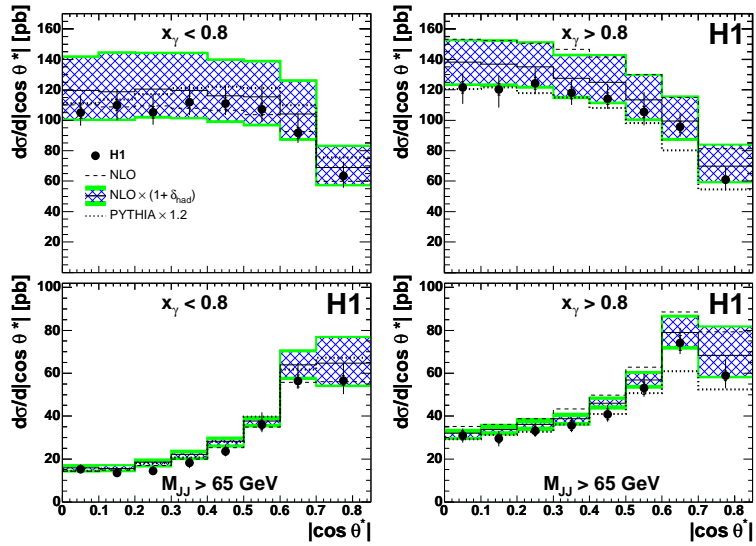,width=4.0in}}
\caption{Dijet production in photoproduction by H1: cross sections 
for resolved-dominated events (left) and direct-dominated events (right)
as a function of dijet centre-of-mass scattering angle.
}\label{Hdij}
\end{figure}
H1 have measured the photoproduction of dijet systems
with improved statistics (Fig.\
\ref{Hdij}).\cite{H1phdij}  
When the dijet mass is above 65 GeV, the distributions in
$|\cos\theta^*|$ show a clear difference between the direct- and
resolved-dominated samples.  In direct processes, the incoming photon
interacts with the proton by coupling to a high-$E_T$ quark line: the
hard exchanged object is thus a virtual quark, with spin one-half.  In
resolved processes, a parton from the partonic substructure of the
photon interacts with a quark or a gluon from the proton, and the
exchanged object is more likely to be a spin-one gluon.  Hence the QCD
dynamics is different: a less steep $|\cos\theta^*|$ distribution
occurs in the direct case owing to the exclusively spin-half propagator.

The data are well described by PYTHIA, in which the QCD dynamics is of
course inbuilt, rescaled by a factor of 1.2; this does not
affect the shape of the distributions.  
An NLO calculation describes a variety of distributions to within 10\%.\cite{Frix}

\begin{figure}[t]
~\\[-3mm]
\centerline{
\mbox{
\psfig{file=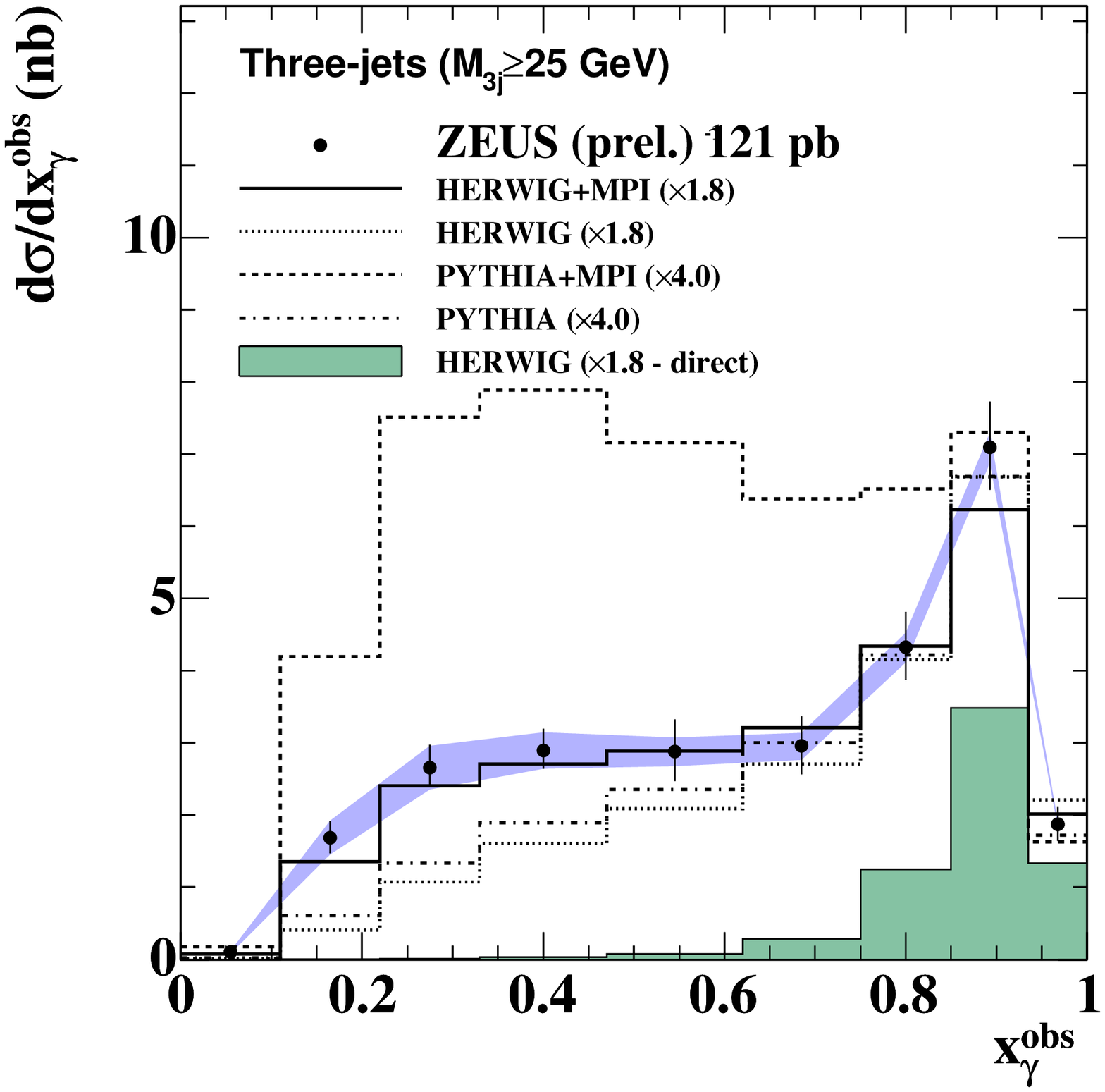,width=2.4in}
\psfig{file=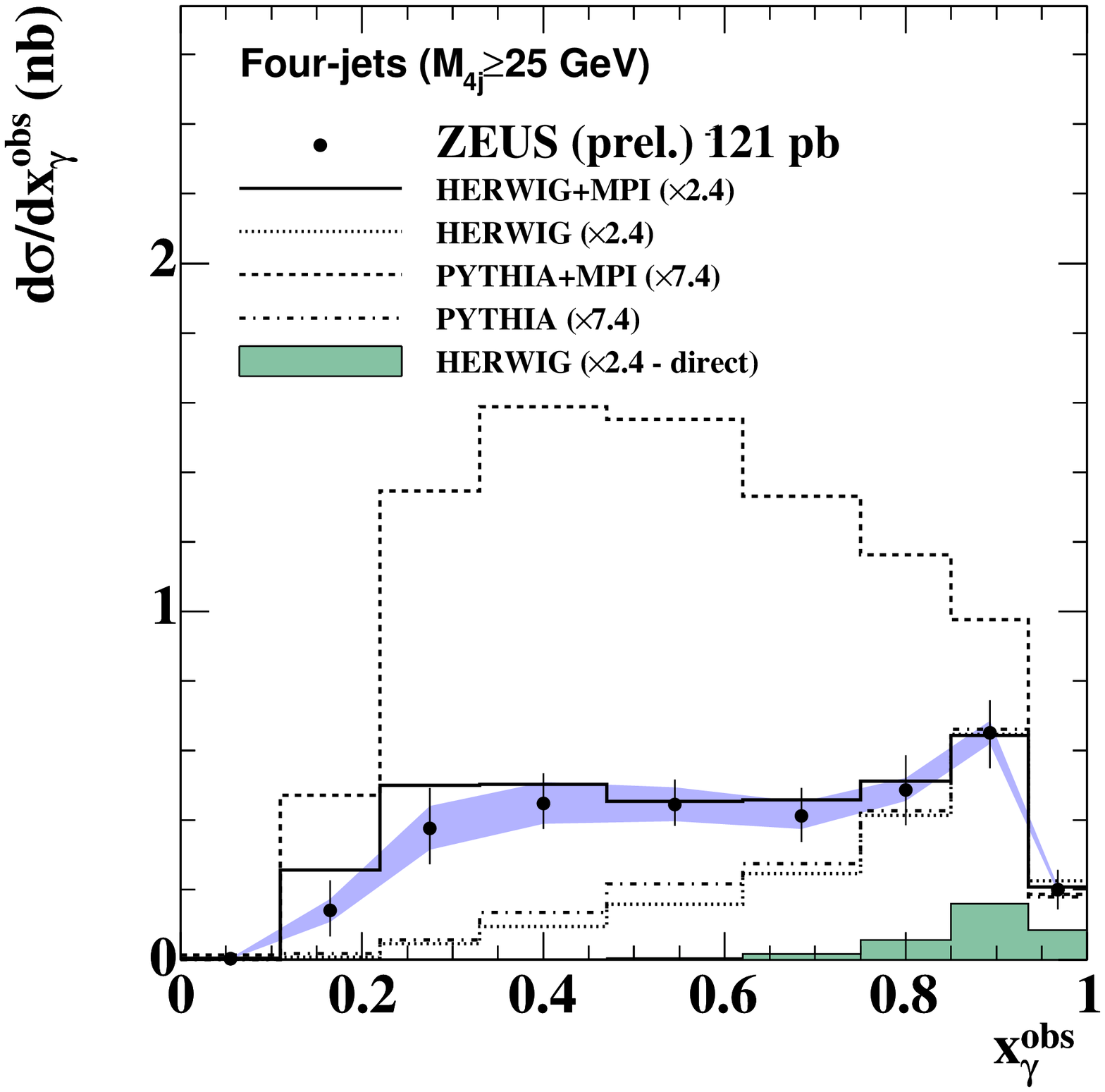,width=2.4in}
}}
\caption{Three and four jet final states in photoproduction in ZEUS:
the fraction of photon momentum carried  by the jets, compared to 
some theoretical models.
  }\label{Zphpmjets}
\end{figure}
Three- and four-jet final states have been measured in photoproduction
at HERA by ZEUS.  The integrated luminosity of 121 pb$^{-1}$
represents over seven times the luminosity of the previous HERA
publication on three jets in photoproduction, while the four-jet
photoproduction cross section has been measured for the first time and
constitutes the highest order process so far studied at HERA. The
events have been studied for masses of the jet system greater than 25
GeV and greater than 50 GeV. A comparison with the PYTHIA and HERWIG
shows that a multiple parton interaction scheme is necessary, but that
tuning to $ep$ data is necessary; this has been done with HERWIG but
not yet with PYTHIA (Fig.\ \ref{Zphpmjets}).  The three-jet cross
sections have been compared to an $O(\alpha\alpha_s^2)$ perturbative
QCD calculation.\cite{KKK} This describes the higher-mass region
reasonably well but underestimates the data for jet system masses
below 50 GeV.

\section{Heavy flavour studies}

The dominant process for heavy flavour production at HERA is
photon-gluon fusion: $\gamma g \to Q \bar{Q}$.  The cross sections
therefore depend strongly on the gluon density in the proton. There
are two main theoretical approaches, namely the zero-mass variable
flavour number scheme (ZM-VFNS) which is suitable for $Q^2$ values much higher
than the quark mass squared, and the fixed-flavour number scheme which
allows for quark masses, but which in present implementations becomes
inaccurate at high $Q^2$ owing to large terms of the form $\log Q^2/M^2$ in
the perturbative expansion.  Combinations of the two schemes have been
devised, referred to simply as VFNS.

\begin{figure}[t]
\centerline{
\mbox{
\hspace*{-2mm}
\psfig{file=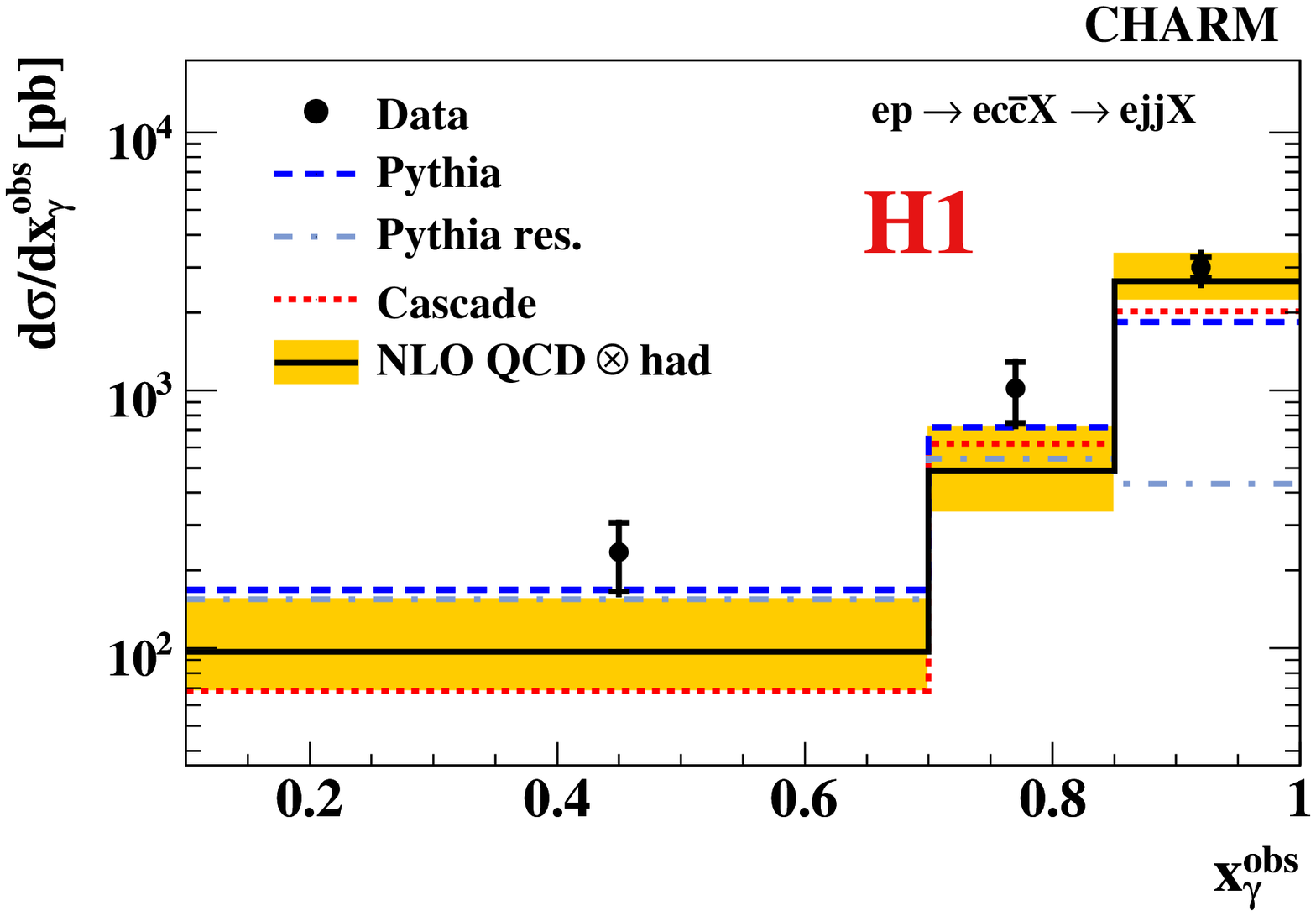,width=2.4in}
\hspace*{2mm}
\psfig{file=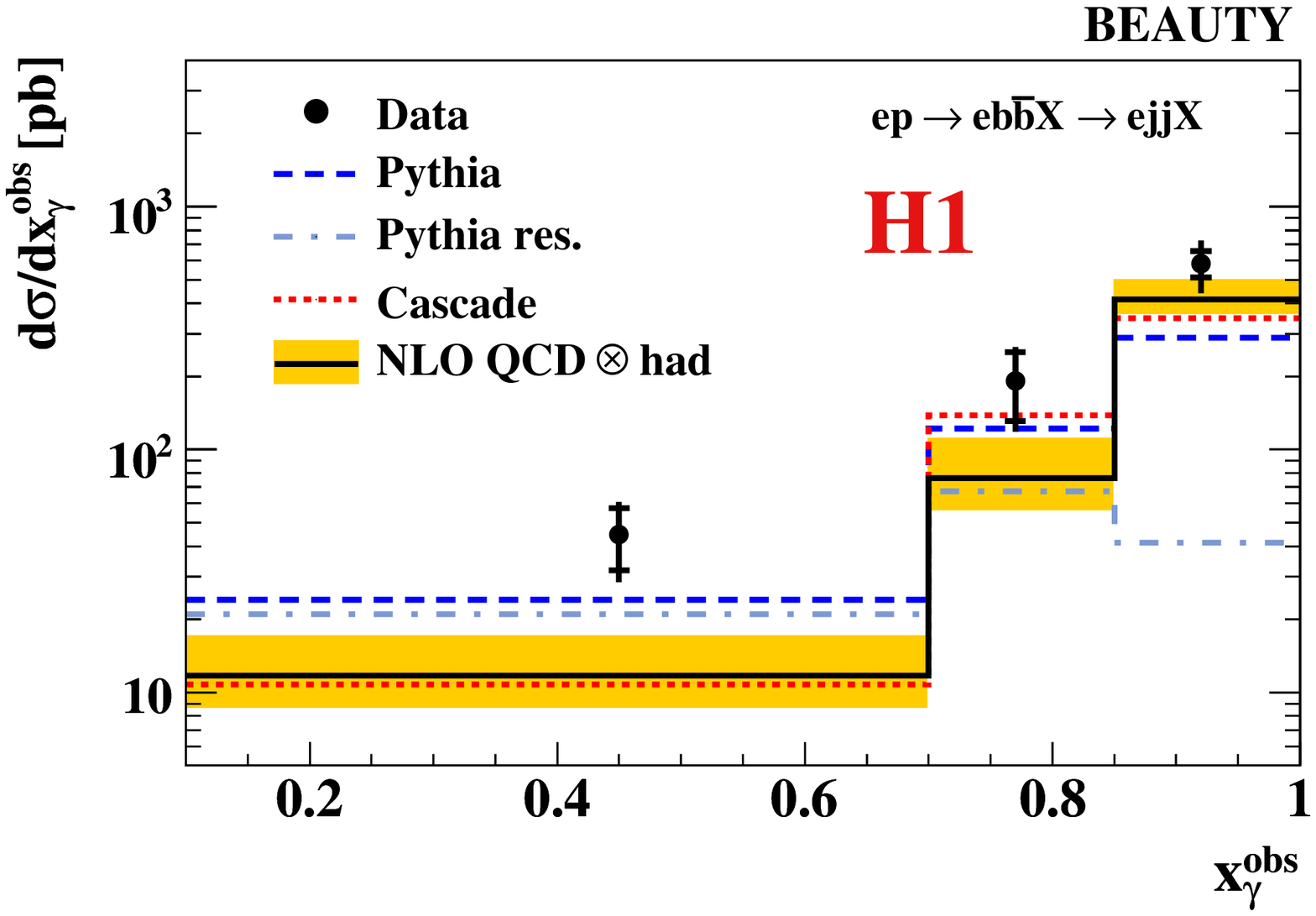,width=2.4in}
}}
\caption{The fraction of the photon momentum taken by 
a dijet system containing charm or beauty. Events are selected with
two or more jets of transverse momentum above 11, 8 GeV respectively
in the central laboratory pseudo-rapidity range $-0.9$ - $1.3$.
}\label{HBCprp}
\end{figure}

An analysis of charm and beauty dijet photoproduction cross sections
has been carried out by H1.\cite{H1bbcc} The fractions of events
containing charm and beauty quarks are determined using a method based
on the impact parameter in the transverse plane of tracks relative to
the primary vertex, as measured by the H1 central vertex detector. A
variety of differential dijet cross sections for charm and beauty 
and their relative contributions to the flavour inclusive dijet
have been calculated, as illustrated in
Fig.\ \ref{HBCprp}). Comparison is made with an NLO QCD
calculation.\cite{Frix0} Taking into account the theoretical
uncertainties, the charm cross sections are consistent with a QCD
calculation at NLO, while the predicted cross
sections for beauty production are somewhat lower than the measured
values in the resolved-dominated region of the distribution.
For direct-dominated events, the respective contributions of charm and 
beauty events to the total dijet cross section should be 4/11 and 1/11,
given by the squares of the charges of the respective quarks,
for dijet masses much greater than the quark mass.  This is
found to hold within experimental uncertainties. 

\begin{figure}[t]
~\\[-8mm]
\centerline{
\mbox{\hspace*{-5mm}
\psfig{file=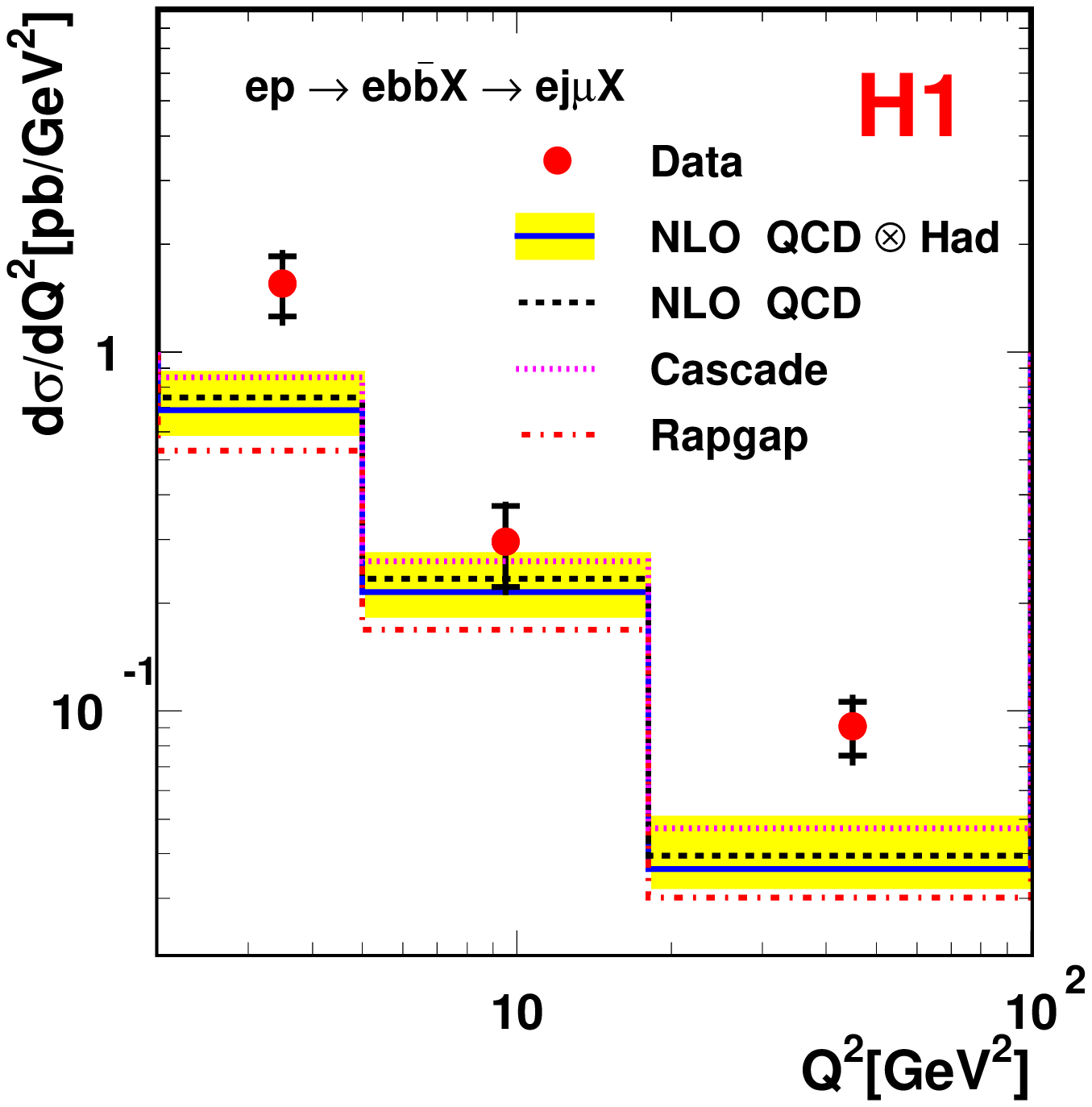,width=2.1in}
\hspace*{-4mm}
\psfig{file=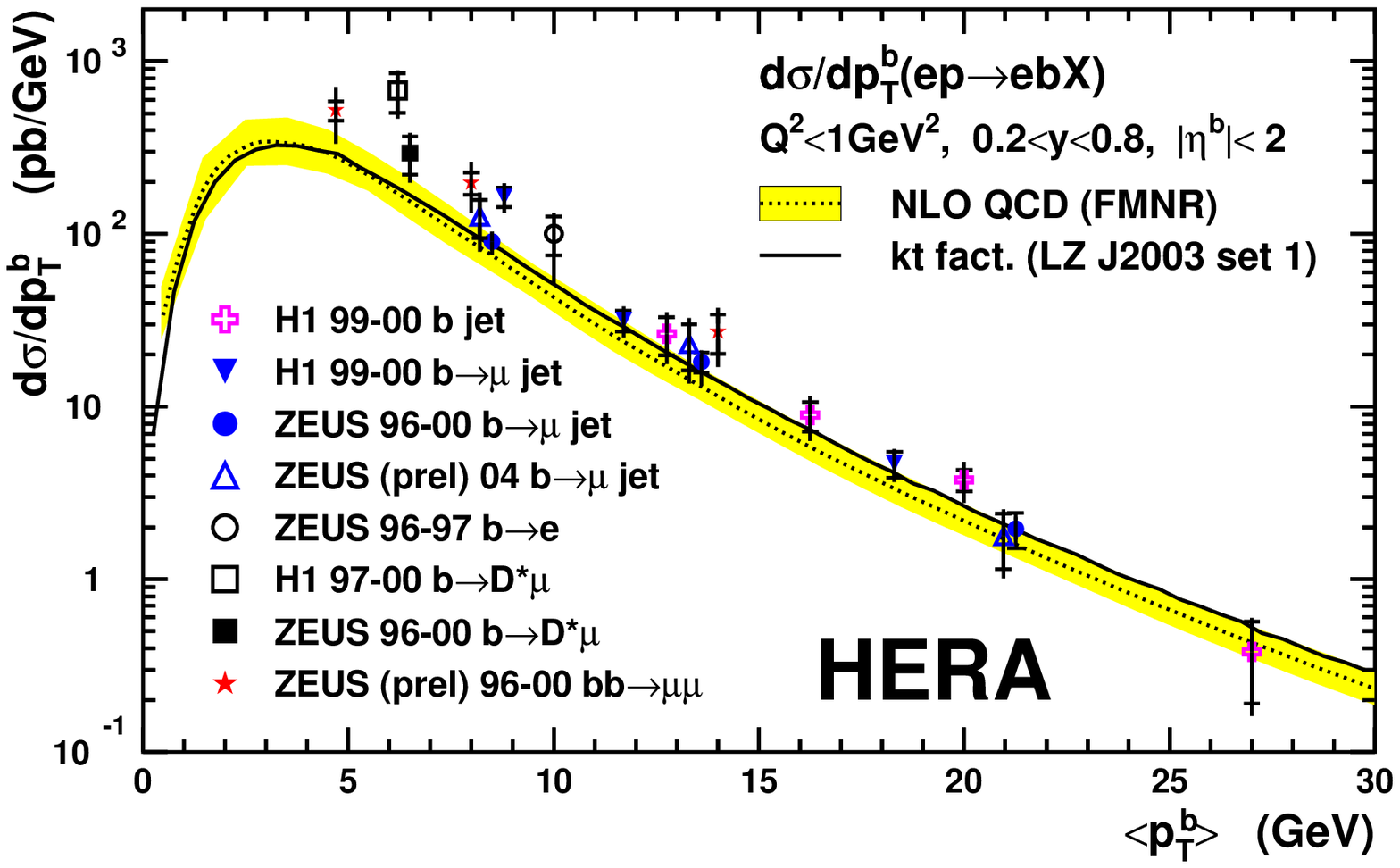,width=3.4in}}
}
\caption{(Left) H1 cross section for $b$ quark production in DIS
as a function of $Q^2$ compared to theoretical models. (Right)
cross section for $b$ quark photoproduction 
in ZEUS (preliminary), with earlier results
and NLO QCD prediction.}\label{ZBmumu}
\end{figure}

These results follow a measurement of the beauty production cross
section in ep collisions, both photoproduction and DIS, in which
events were selected by requiring the presence of jets and muons in
the final state.\cite{H1Bmuj} Both the long lifetime and the large
mass of b-flavoured hadrons were exploited to identify events
containing beauty quarks. The differential cross sections in
photoproduction, and in deep inelastic scattering, with $2 < Q^2 < 100$
GeV$^2$ were compared with perturbative LO and NLO QCD calculations.
As evidenced in Fig.\ \ref{ZBmumu}, the
predictions are found to be somewhat lower than the data.

Beauty photoproduction with events in which two muons are observed in
the final state has been measured with the ZEUS detector at HERA
(Fig.\ \ref{ZBmumu}, right). 
A low $p_T$ threshold for muon identification, in combination with
the large rapidity coverage of the ZEUS muon system, gives access to
essentially the full phase space for beauty production. The dimuon
selection suppresses backgrounds from charm and light flavour
production. The total cross section for beauty production has been
measured and differential cross sections for $b$ quark production have
been extracted on the basis of our knowledge of $b$ fragmentation. A
comparison to NLO QCD predictions\cite{Frix0}
and to other ZEUS measurements is shown
in Fig.\ \ref{ZBmumu}.  Although no single measurement is
significantly different from theory, taking theoretical and
experimental uncertainties into account, there is a consistent trend for
theory to be low at lower jet momenta, although high and low values of $Q^2$
do not appear to differ in this respect. 

\begin{figure}[t]
~\\[-6mm]
\centerline{
\mbox{
\psfig{file=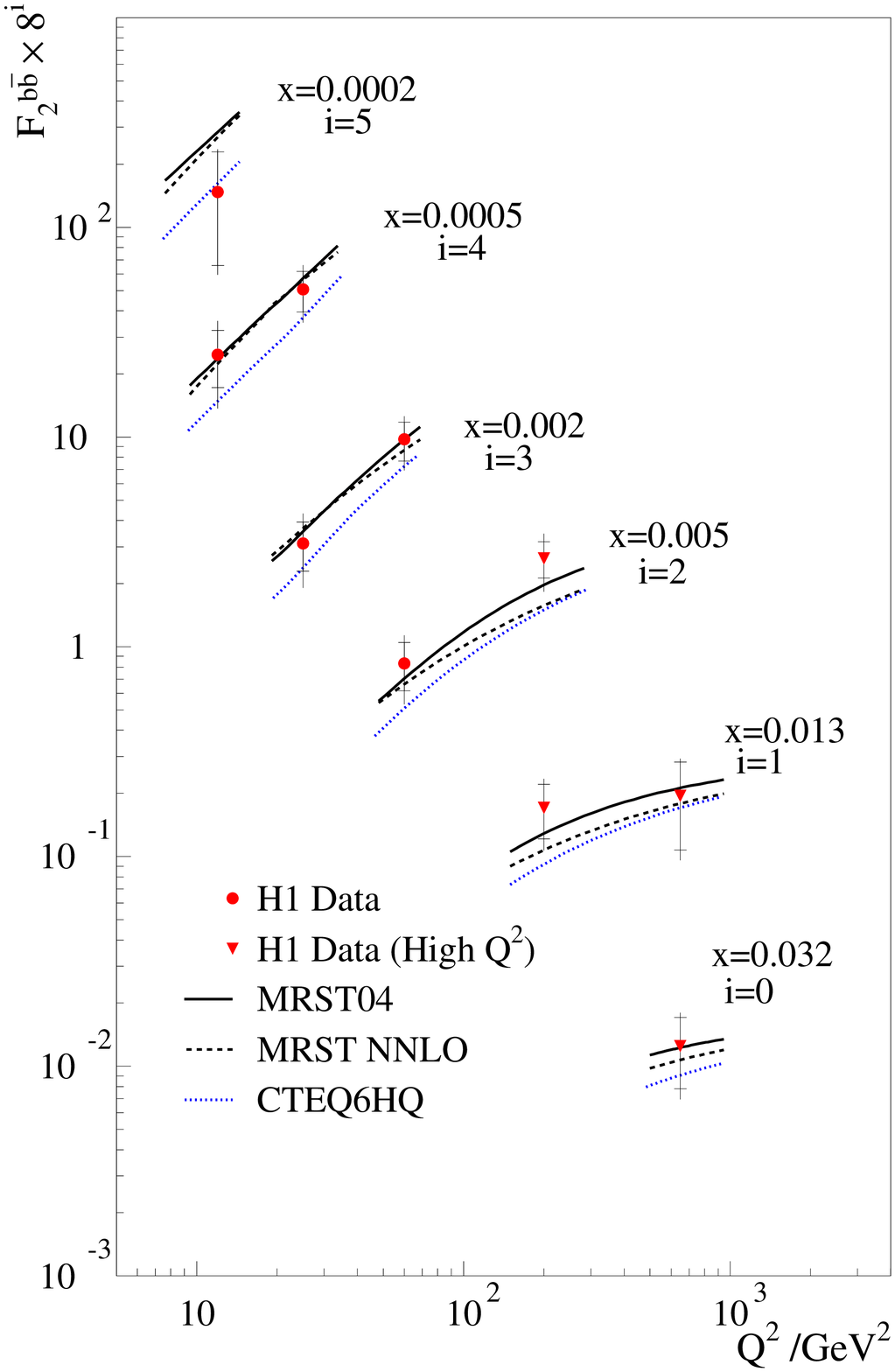,width=2.5in}
\psfig{file=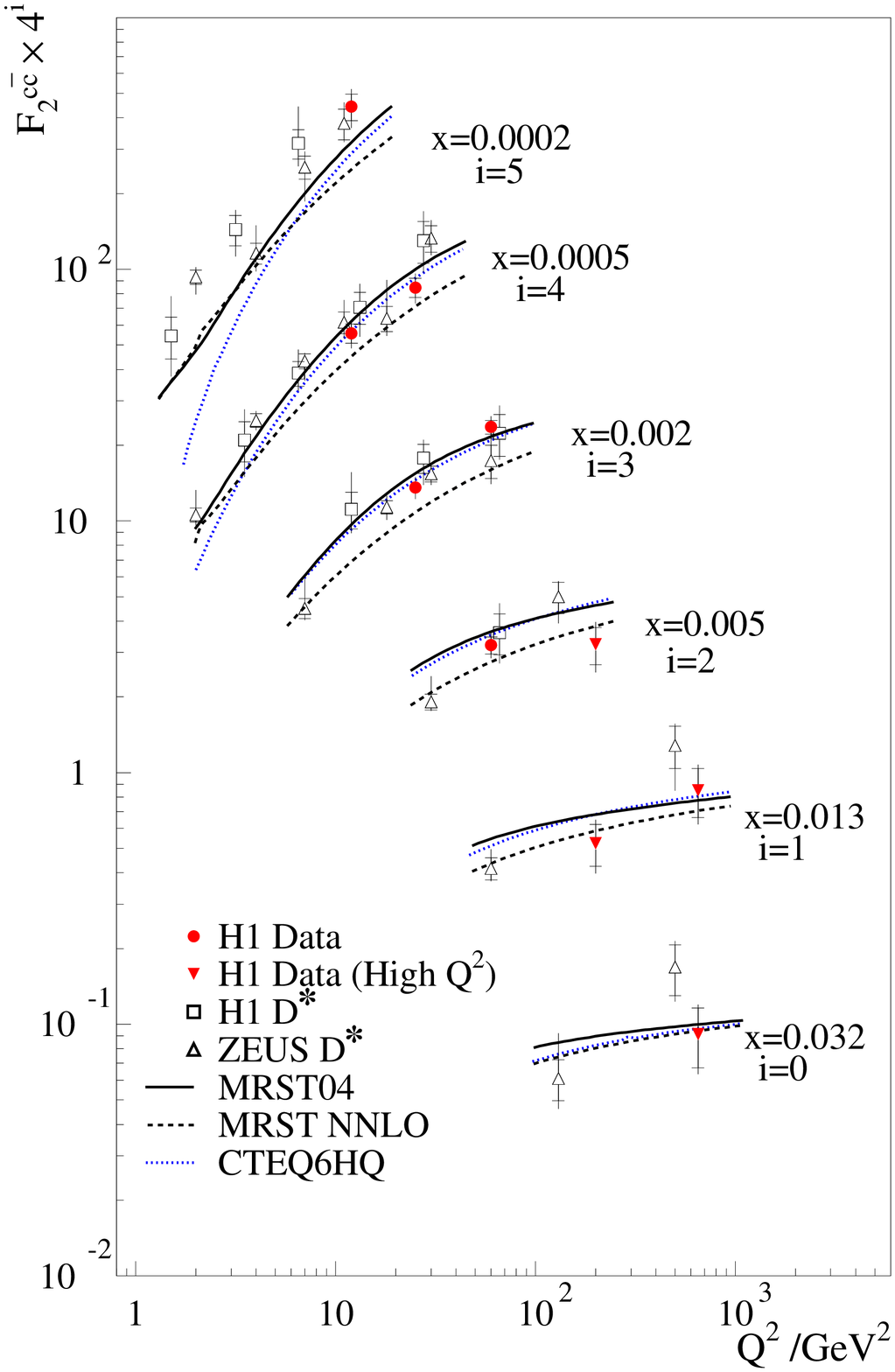,width=2.5in}
}}
\\[-4mm]
\caption{Measurements of the proton structure function $F_2$ by H1 for 
processes involving production of $b$ and $c$ quarks in the final 
state, compared  to models with different proton PDFs. }
\label{H1bbcc}
\end{figure}
  H1 have presented new results for the proton structure
function measured for the production of final states involving $c$ and
$b$ quarks (Fig.~\ref{H1bbcc})\cite{H1bbccDIS}.  Beauty and
charm comprise respectively 0.4 - 3\% and $\approx$24\% of the total cross section
over this kinematic range.  These are the first
such measurements for $b$ quarks and were made exploiting the H1 silicon
vertex detector. The slopes of the distributions with $Q^2$
at given $x$ (scaling variations) arise from perturbative-QCD effects
on the PDF's and are in accord with predictions.

The H1 data are compared with several VFNS-based theoretical
approaches incorporating different proton PDFs.  The agreement is
reasonable for both charm and beauty, but the present experimental and
theoretical uncertainties are such that the different models cannot
yet be distinguished.

\section{Diffractive production}
\begin{figure}[t]
\centerline{
\hspace*{10mm}\psfig{file=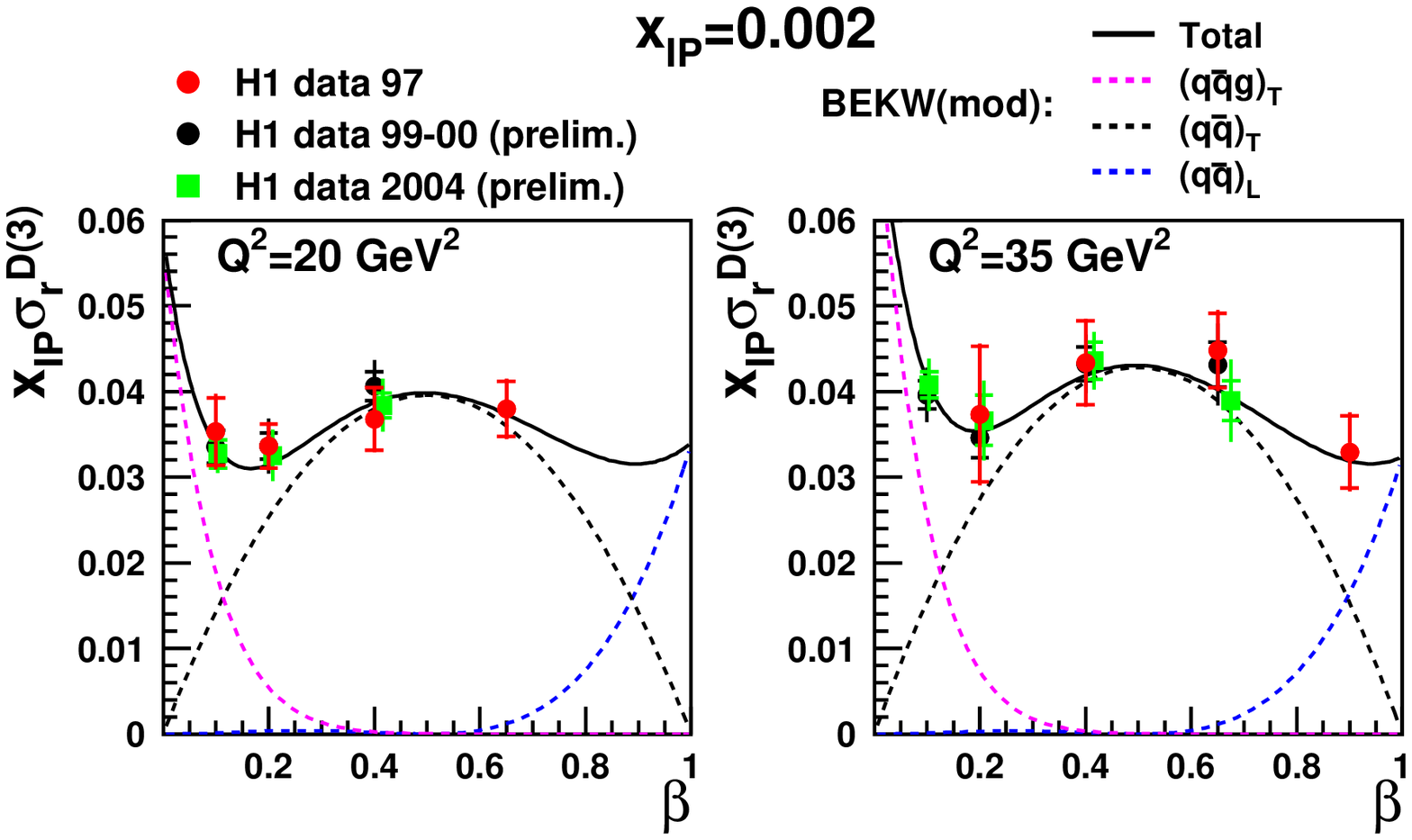,width=4in}
}
\caption{Measurents of
the reduced cross section $\sigma_r^{D(3)}$ for the diffractive
process $ep\to eXY$ (where $Y$ denotes the proton or a possible
low-mass diffractive excitation) for $12 \leq Q^2 \leq 120$
GeV$^2$ and $|t| < 1.0$ GeV$^2$. }
\label{H1diff}
\end{figure}
Some recent results on diffractive production in DIS are illustrated in
Fig.\ \ref{H1diff}.  The quantity $\sigma_r^{D(3)}$ measures the
inclusive differential cross section and is a good approximation to
the diffractive proton structure function.  The quantities $x_{I\!\!P}$
and $\beta$ approximate respectively to the fraction of the proton
momentum carried by the colourless exchanged ``pomeron'', and to the
fraction of the latter's momentum carried by the struck quark.

The figure shows results obtained by H1 at $x_{I\!\!P} =
0.002$ for two values of $Q^2$.  There is good agreement
with the predictions of a model in which the ``pomeron'' is modelled
as two gluons.\cite{BEKW} The results are also in good agreement with
earlier H1 and ZEUS diffractive measurements.\cite{HZdiff}

\section{Conclusions}
The results presented here have of necessity been selective,
chosen to illustrate the range and depth of QCD  
studies that are being carried out at HERA.  Much
progress has been made over recent years,  in the type of studies that
can be performed, the precision achieved, and in theoretical understanding.
HERA will complete its data-taking phase in mid-2007. The final analysis
of its results will provide a unique corpus of knowledge that will not be
overtaken for the foreseeable future.

\section*{Acknowledgments} 
I should like to thank colleagues in ZEUS and H1, in particular
Olaf Behnke and Massimo Corradi, for helpful conversations and comments.

\end{document}